\documentstyle[prd,aps,epsfig,eqsecnum,amssymb,amsmath,floats]{revtex}

\begin{document}
\draft
\preprint{Draft Nb. 0}
\title{On the parametrization of atmospheric muon angular flux underwater}
\author{S. I. Klimushin, E. V. Bugaev, I. A. Sokalski}
\address{
Institute for Nuclear Research, Russian Academy of Sciences, 60th October Anniversary prospect 
7a, Moscow 117312, Russia
}
\date{\today}
\maketitle
\begin{abstract}
The analytical expression for angular integral flux of atmospheric muons in matter with the explicit
relation of its parameters with those of the sea level spectrum is obtained.
The fitting formula for the sea level muon spectrum at different zenith angles for spherical atmosphere
is proposed. The concrete calculations for pure water are presented.
Fluctuations of muon energy losses are taken into account by means of parametrized correction factor
calculated using survival probabilities resulted from Monte Carlo simulations.
Parametrizations of all continuous energy losses are obtained with using the most recent expressions 
for muon interaction cross-sections.
The corresponding parametrization errors and field of method application are
comprehensively discussed. The proposed formulae could be useful primarily for experimentalists processing
data of arrays located deep under water or under ice. 
\end{abstract}

\widetext
\section{INTRODUCTION}
\label{sec:int}
The last several years have been marked by starting of fullscale data taking of large neutrino
and muon telescopes located at lake Baikal (NT-200~\cite{NT-200}) and in deep polar ice (AMANDA~\cite{amanda}). 
Two underwater telescopes
ANTARES~\cite{antares} and NESTOR~\cite{nestor} assuming installation at bigger depths are intensively 
constructed in the Mediterranean. The possibility of deployment of telescopes with huge detecting 
volumes up to 1 km$^3$ (KM3) is also under wide investigation.
 
So, the knowledge of expected angular distribution of integral flux of atmospheric muons deep underwater  
is of interest not only for cosmic ray physics 
but also for estimation of possible background for neutrino detection and at last for a test 
of the correctness of underwater telescope data interpretation by using the natural flux of atmospheric muons as calibration source.    
The last item frequently implies the estimation with an appropriate accuracy (e.g., better than 5$\,\%$
for a given sea level spectrum) the 
underwater integral muon flux for various sets of depths, cut-off energies and angular bins 
especially for telescopes of big spacial dimensions (e.g., AMANDA, KM3).

The existing methods of calculations of muon propagation through thick layers of matter 
(basically for a standard rock) are largely presented in the literature.
There are two completely different approaches in such calculations. In first (historically) works on
the subject of muon propagation the integro-differential kinetic equation for muon flux had been formulated and
approximately solved using semianalytical methods. We mention here one of the pioneer works of this approach:
Ref.~\cite{ZM62}, and one the most recent papers: Ref.~\cite{NSB94} (this latter paper contains also the large bibliography).
Works of the second approach use Monte Carlo (MC) technique for propagation studies: the main element in
such works is simulation of sequences of free flights and interactions during muon passage through a medium. 
One of the most early works using this method is Ref.~\cite{HPW63}, the most recent one is the work of the present authors
Ref.~\cite{MUM1}, containing also many bibliographic references.

Up to now the presentation of the results of these calculations both for parent muon sea level spectra (especially for angular
dependence taking into account the sphericity of atmosphere) and for underwater angular flux was not quite convenient when applying 
them to concrete underwater arrays.
In addition, a part of numerical results is available only in data tables (often insufficient for accurate interpolation) 
and figures.
The possibility of direct implementation of MC-methods depends on the availability of corresponding codes
and usually assumes rather long computations and accurate choice of the grid for simulation parameters to avoid big 
systematic errors.   

Therefore, the necessity of analytical expressions both for underwater muon integral flux and underwater muon differential 
spectrum is still actual. Besides, the possibility of reconstructing the parameters of a sea level
spectrum by fitting measured underwater flux in the case of their direct relation looks rather attractive. 
The only (to our knowledge) existing analytical description of muon underwater angular flux published 
in Ref.~\cite{Okada} is rather comprehensive but its relation to widely used parametrizations 
of sea level spectrum is unclear.   
  
In this paper we present rather simple method allowing to calculate analytically 
the angular distribution of integral muon flux
deep under water for cut off energies (1--$10^4)\,$GeV and slant depths of (1--16)$\,$km 
for conventional ($\pi,\,K$) sea level spectra of atmospheric muons having widely used parametrization by means of 5 parameters. 
The fluctuations of muon energy losses are taken into account. 

The plan of the paper is as follows. In Sec.~\ref{sec:Method} the idea of the method used for the derivation of 
the analytical expression for an underwater integral muon flux is described.
In Sec.~\ref{sec:Sealm} the common analytical description of widely used sea level muon angular spectra is introduced.
In Sec.~\ref{sec:Mupr} the parametrization of muon continuous energy losses is given. In the same section the calculation of
survival probabilities needed for a taking into account muon energy loss fluctuations is discussed.
In Sec.~\ref{sec:Underw} the concept of an correction factor is introduced. The results of numerical calculations and
analytical parametrization of this factor are given. At the end of this section the final analytical expression for the
angular integral underwater flux is derived. 
The main conclusions of the work are given in Sec.~\ref{sec:Concl}. Some mathematical details are presented in Appendices~\ref{app:a1}
and~\ref{app:a2}.
         
\section{Method}
\label{sec:Method}
The integral flux of muons in matter with energies above cut-off $E_f$
expected at slant depth $R$ at a zenith angle of $\theta$ taking into account 
{\it fluctuating} character of muon losses is conventionally described by
\begin{equation}\label{Int_fl}
I_{fl}(\geq E_f,R,\theta)=\int_0^{\infty}{P(E_0,R,\geq E_f)D(E_0,\theta)\,dE_0},
\end{equation}
where $P(E_0,R,\geq E_f)$ is the probability that a muon, having starting energy $E_0$,
after passing of path $R$ will survive with final energy above cut-off $E_f$
and $D(E_0,\theta)=dN(E_0,\theta)/dE_0$ is a sea level differential angular spectrum.
In a case of a flat surface the slant depth is expressed by $R=h/\cos\theta$, where $h$
is a vertical depth below the surface.

With the assumption of {\it continuous} energy loss rate of muon in matter, $L(E)=-dE/dx=a(E)+b(E)E$, the 
integral flux $I_{cl}$ is derived from the equation 
\begin{equation}\label{Int_cont}
I_{cl}(\geq E_f,R,\theta)=\int_{E_s}^{\infty}{D(E_0,\theta)\,dE_0},
\end{equation}
where the low limit of the integral, $E_s$, is the value of starting energy $E_0$ which results, 
after passing of path $R$, in the final energy $E_f$. This value is derived from the solution of the integral equation
\begin{equation}\label{Ra}
-\int_{E_s}^{E_f} \frac{dE}{L(E)}=R.
\end{equation}
The method proposed in this work for an obtaining the approximate {\it analytical} expression for the 
real integral muon flux allowing for loss fluctuations consists in the following.
Firstly, we calculate numerically $I_{fl}$ and $I_{cl}$ and the {\it correction factor} defined by the ratio 
\begin{equation}\label{Cf}
C_{f}(\geq E_f,R,\theta)=\frac {I_{cl}(\geq E_f,R,\theta)} {I_{fl}(\geq E_f,R,\theta)}.  
\end{equation}
In principle, this factor can be calculated using known codes for muon propagation through matter. 
In this work we apply for this aim the MUM code
described in previous work of the present authors~\cite{MUM1}. 
Further, we approximate the muon energy loss function $L(E)$ by a linear energy dependence,
\begin{equation}\label{Leff}
                               L(E)=\alpha + \beta E
\end{equation}
($\alpha$ and $\beta$ are independent on energy) and parametrize the sea level muon spectrum by dependencies
of the type
$$
D(E_0) \propto \frac {E_{0}^{-\gamma}} {1+E_{0}/\varepsilon_{0}}.
$$
Using these approximations, the integral flux in matter defined by Eq.~(\ref{Int_cont}) can be expressed analytically. 
Below in this paper we will designate this integral flux by $F_{cl}$ (to
distinguish it from $I_{cl}$ calculated numerically).

The real integral flux allowing for energy loss fluctuations 
can now be obtained using the correction factor introduced in Eq.~(\ref{Cf}). As it is shown below, the dependencies of this factor on
its arguments are rather smooth and can be easily parametrized, so the final result for the real integral muon flux in a matter is also
given by the analytical expression:
\begin{equation}\label{For}
F_{fl}(\geq E_f,R,\theta)=\frac {F_{cl}(\geq E_f,R,\theta)} {C_{f}(\geq E_f,R,\theta)}.  
\end{equation}
%
\protect\section{Sea level spectra of atmospheric muons}
\label{sec:Sealm}
From the large collection of conventional ($\pi,~K$) angular spectra of atmospheric muons at sea level $D(E_0,\theta)$
proposed in literature~\cite{NaumovPHR,Sineg,Lipari,Agrawal,Dar,Butk,VZK,Gaisser,LVD1,MACRO}, for the numerical calculations of~(\ref{Int_fl}) and~(\ref{Int_cont}) we have chosen 
as basic the spectrum from work of Ref.~\cite{Sineg}, which had been obtained using calculations described in Refs.~\cite{NaumovPHR,NSS}. 
The reasons are the following:
\begin{itemize}
\item[(i)] the energy range of its validity is up to $10^9$ GeV allowing the calculation of underwater 
flux up to cut-off energy greater than 10 TeV for slant depths as high as (20--30)$\,$km; 
\item[(ii)] access to data tables (given to us by authors of work~~\cite{Sineg}) detailed enough to perform splines via energy and zenith angle 
         with acuracy $<0.5\,\%$;    
\item[(iii)] high level of the agreement of this spectrum with full set of existing experimental results.
\end{itemize}
This spectrum results from the computations based on nuclear cascade model of Refs.~\cite{NCascade,NSS}
(see also Ref~\cite{NaumovPHR}) and semiempirical model for primary spectrum proposed in Ref.~\cite{Nikolsky}.\\          
Following Ref.~\cite{Nprivate}, for the description of basic spectrum  
for the energy range $E_0\leq$1.53 TeV we have used values defined by $D(E_0,\theta)=D(E_0,0^\circ)A(E_0,\theta)$ with vertical
spectrum $D(E_0,0^\circ)$ taken from Ref.~\cite{NaumovPHR} and angular distribution $A(E_0,\theta)=D(E_0,\theta)/D(E_0,0^\circ)$
from data tables based on calculations of Ref.~\cite{Sineg}. For energies above 1.53 TeV we have used values of $D(E_0,\theta)$ according to splines 
computed via the same data tables.  
For the spectrum composed by this way the abbreviation ``NSS'' (following names of authors of Ref.~\cite{NSS}) is used below.

The most convenient analytical form of $D(E_0,\theta)$ has been proposed in Refs.~\cite{Bugaev70,Volkova75} 
and became almost conventional after work~\cite{Gaisser}. It is as follows:  
\begin{equation}\label{Sealds}
\frac{dN(E_0,\theta)}{dE_0}=D(E_{0},\theta)=A_{0}\left(\frac{1}{1+E_{0}/E^{cr}_{0_\pi}(\theta)}+
\frac{B_0}{1+E_{0}/E^{cr}_{0_K}(\theta)}\right)E_0^{-\gamma}.
\end{equation}
Here, $\gamma$ is a spectral index and the functions $E^{cr}_{0_{\pi,K}}(\theta)$ are given by the relation 
\begin{equation}\label{cren}
E^{cr}_{0_{\pi,K}}(\theta)=E^{cr}_{0_{\pi,K}}(0^\circ)/\cos\theta^*.
\end{equation}
These functions have approximate sense of critical energies of pions and kaons for given zenith angle and
$E^{cr}_{0_{\pi,K}}(0^\circ)$ are those for vertical direction. The value $\cos\theta^*$ has sense of
effective cosine taking into account the deviation of real atmosphere from the flat one.
Thus, sea level spectrum may be expressed by 5 parameters ($A_0,B_0,E^{cr}_{0_\pi}(0^\circ),E^{cr}_{0_K}(0^\circ),\gamma$)
each of them having the physical meaning. 

As it is shown in Appendix~\ref{app:a2}, just parametrization~(\ref{Sealds}) is very convenient for a derivation of an analytical 
expression for integral flux in matter. 
To check the validity of this analytical expression for the large range of zenith angles and cut-off energies
we have performed the fit of original NSS spectrum by 5 parameter expression~(\ref{Sealds}) for vertical direction.
This fit was done within ranges $E_0$=(0.2--200)$\,$TeV.

When checking the values of fit spectrum for $\cos\theta$=(0.05--1.0) 
we realized that the standard description of effective cosine (with geometry of spherical
atmosphere and with definite value of effective height of muon generation) is not enough
and one should introduce an additional correction $S(\theta)$ leading to (10--20)$\,\%$
increase of effective cosine value for $\cos\theta<$~0.1. The reason of an appearing of this correction is that the concept of an
effective generation height is approximate one. It fails at large zenith angles where the real geometrical size of the generation region
becomes very large.

\begin{figure}[!t]
\center{\mbox{\epsfig{file=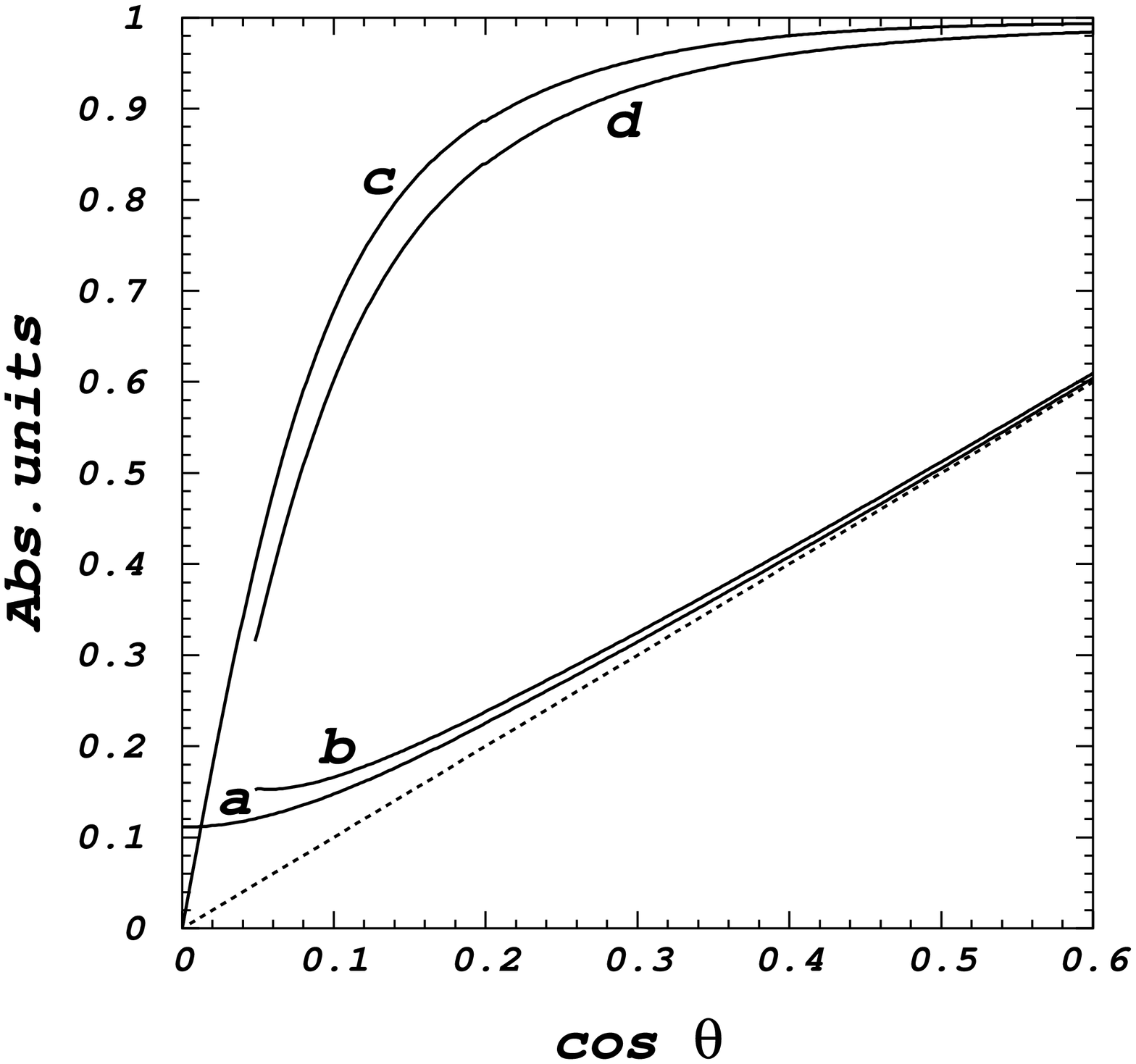,height=9cm}}}
\protect\caption{
Effective cosine as a function of zenith angle.
Curve (a) is geometrical effective cosine $\cos\theta^*$ given 
by Eq.~(\protect\ref{Efcos1}) of Appendix~\protect\ref{app:a1}.
Curve (b) is effective cosine $\cos\theta^{**}$ with the correction and is given by Eq.~(\protect\ref{Efcos}).
Curves (c) and (d) represent the ratio $\cos\theta/\cos\theta^*$ and $\cos\theta/\cos\theta^{**}$, correspondingly. 
\label{fig:efcos}}
\end{figure}

The resulting fit of angular sea level spectrum in units of~(cm$^{-2}$s$^{-1}$sr$^{-1}$Ge$V^{-1}$) is given by
\begin{equation}\label{BKfit}
D(E_0,\theta)=0.175\left(\frac{1}{\displaystyle 1+\frac{E_0\cos\theta^{**}}
{\displaystyle 103}}+\frac{0.037}{\displaystyle 1+\frac{E_0\cos\theta^{**}}
{\displaystyle 810}}\right)E_0^{-2.72},
\end{equation}
with modified effective cosine expressed by
\begin{equation}\label{Efcos}
\cos\theta^{**} =S(\theta)\cos\theta^{*},  
\end{equation}
where $\cos\theta^*$ is derived from spherical atmosphere geometry and is given by formulae of Appendix~\ref{app:a1}
and $S(\theta)$ is the correction which is given for $\sec\theta\leq20$ by
\begin{equation}\label{Scos}
S(\theta)=0.986+0.014\sec\theta.  
\end{equation}
The dependencies of both modified and geometrical effective cosines as function of zenith angle are presented  
in Fig.~\ref{fig:efcos}. Note that for $\cos\theta>$~0.4 the influence of the curvature of real atmosphere is less
than 4~$\%$ but for $\cos\theta<$~0.1 it is greater than 40~$\%$. 

We should note that our expression~(\ref{Sealds}) for the sea level muon spectrum does not contain a contrubution from
atmospheric prompt muons. According to the most recent calculations based on perturbative QCD, this contribution becomes 
essential only at $E_0 > 10^6$ GeV~\cite{Gelm}. The predictions of nonperturbative models (see, e.g.,~\cite{bugPM})   
are slightly more optimistic. We plan to generalize our approach and include this contribution in our following paper.
Incidentally, inclusion of prompt muons should be done in parallel with taking into account the steepening of the sea
level muon spectrum due to the knee in the primary cosmic ray spectrum.
 
\begin{figure}[!t]
\center{\mbox{\epsfig{file=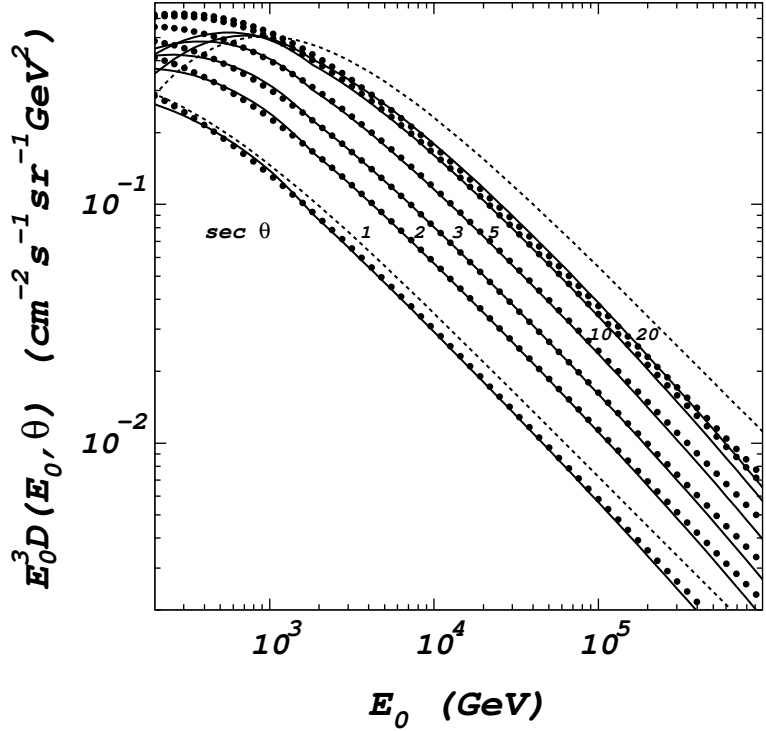,height=10.1cm}}}
\protect\caption{
Differential spectra of conventional muons at sea level for six zenith angles, sec$\,\theta$: 1, 2, 3, 5, 10, and 20, from bottom to top.  
Curve labels correspond to values of sec$\,\theta$.  
Solid curves -- NSS spectrum, dotted curves -- spectrum defined by Eq.~(\protect\ref{BKfit}), 
dashed curves -- spectrum of Ref.~\protect\cite{VZK}
with an angular distribution from Ref.~\protect\cite{Volkova75} shown only for two values of sec$\,\theta$ 1 and 20.  
\label{fig:Seal}}
\end{figure}

Fig.~\ref{fig:Seal} illustrates the limits of applicability of angular spectrum given by Eq.~(\ref{BKfit}), for energy and zenith angle
variables.
The energy region, inside which the deviation from parent NSS spectrum is less than 5~$\%$, is shifted from    
(0.3--200)$\,$TeV for $\cos\theta$=1.0 to (1.5--300)$\,$TeV for $\cos\theta$=0.05.  
The sea level spectrum given by~(\ref{BKfit}) is valid only below the knee ($E_{0}\sim\,$300 TeV) of primary cosmic ray spectrum.
Below we will check the influence of the knee to the value of underwater flux by comparison with
calculations resulted from NSS sea level spectrum which takes into account the knee.

\begin{figure}[!t]
\center{\mbox{\epsfig{file=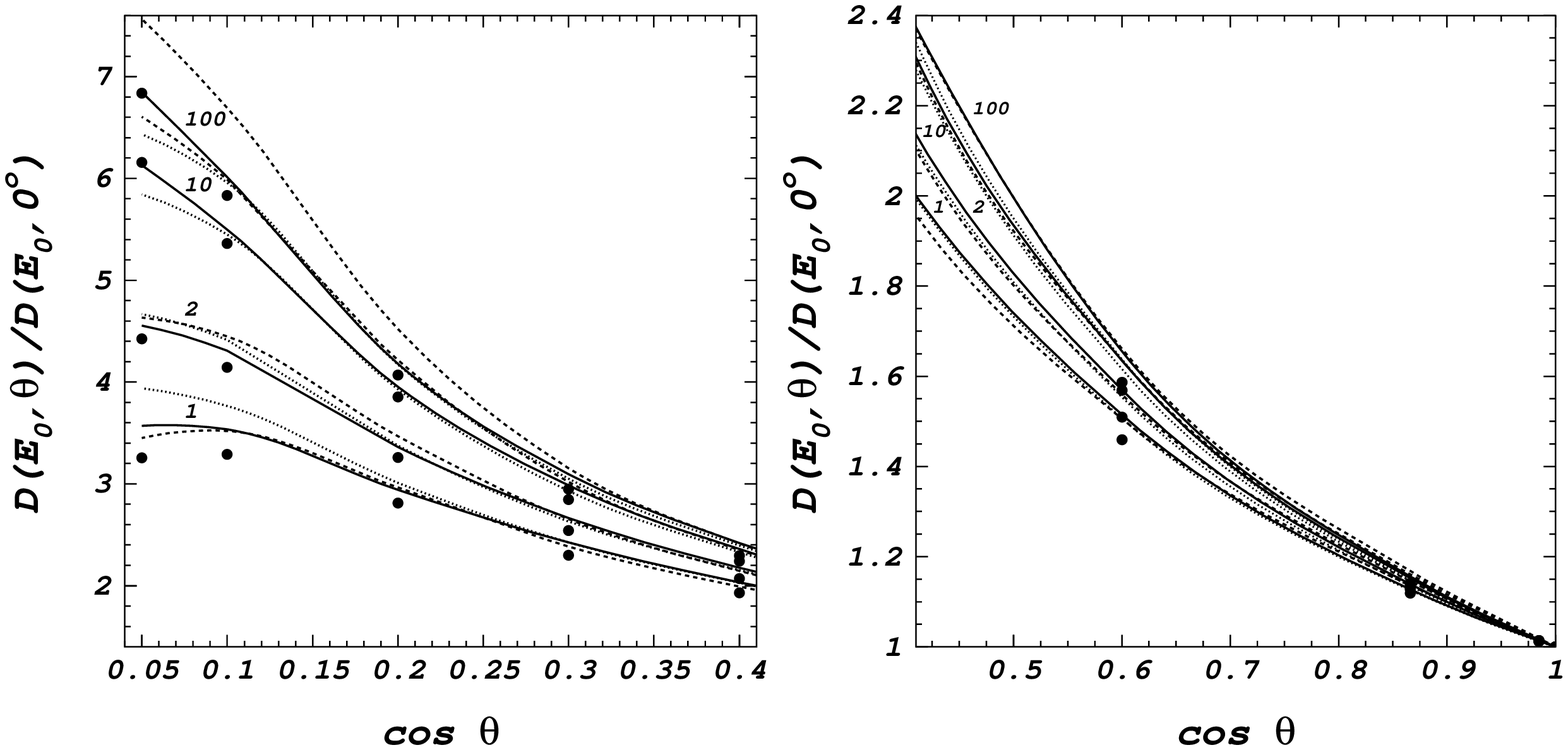,height=10.cm,width=18.cm}}}
\protect\caption{
Sea level muon angular distribution defined by the ratio of differential flux at $cos\theta$ to that at
vertical direction. The disributions are shown for four values of energy $E_0$ (TeV): 1, 2, 10, and 100, from bottom to top. 
Curve labels correspond to values of $E_0$.  
The regions of small (right picture) and big (left picture) zenith angles are enlarged for more details. 
Solid curves -- NSS spectrum, dotted curves -- spectrum defined by Eq.~(\protect\ref{BKfit}), dashed curves -- are taken from 
Ref.~\protect\cite{Volkova75} (for $\pi + 15\,\%~K$), and closed circles -- spectrum of Ref.~\protect\cite{Butk}. 
\label{fig:Ang}}
\end{figure}

Despite the parametrization~(\ref{BKfit}) was done using absolute values of $D(E_0,\theta)$, it reveals rather small deviation  
of the angular distribution $A(E_0,\theta)=D(E_0,\theta)/D(E_0,0^\circ)$ (shown in Fig.~\ref{fig:Ang}) from that of 
parent NSS spectrum.
Note rather good consistence of angular distribution of NSS spectrum with that 
of Ref.~\cite{Butk} for $\cos\theta<\,$0.2 
and differences as big as 10~$\%$ from that of Ref.~\cite{Volkova75} for energies $E_0>\,$10 TeV and $\cos\theta<\,$0.2.
The angular distributions for $E_0>\,$100 TeV are not shown in Fig.~\ref{fig:Ang} because of their coincidence 
with those for $E_0$=100 TeV. 

The widely used vertical sea level spectra which can be parametrized by~(\ref{Sealds}) are presented in 
Table~\ref{tab:seal_all}.
\begin{table}[htb]
\protect\caption{ Parameters of the fitting formula~(\protect\ref{Sealds}) for some selected vertical sea level
muon spectra used in this paper. The values of critical energies are in (GeV) and values of other parameters 
allow the calculation of spectrum in units of (cm$^{-2}$s$^{-1}$sr$^{-1}$Ge$V^{-1}$). Corresponding energy ranges 
of fit validity and references are mentioned.
\label{tab:seal_all}}
\center{\begin{tabular}{llllllll}
Name & $A_0$ & $B_0$ & $E^{cr}_{0_\pi}(0^\circ)$ & $E^{cr}_{0_K}(0^\circ)$ &~~~$\gamma$ & $E_0$ (TeV) & Ref\\\hline
VZK      & 0.1258 & 0.0588 & 100     & 650     & 2.65 & 0.1 -- 100  & \cite{VZK}\\
Gaisser  & 0.14   & 0.054  & 104.545 & 772.727 & 2.70 & 0.03 -- 100 & \cite{Gaisser}\\
Present work& 0.175  & 0.037  & 103     & 810     & 2.72 & 0.3 -- 250  & Eq.~(\ref{BKfit})\\
LVD      & 0.2576 & 0.054  & 104.545 & 772.727 & 2.77 & 2.0 -- 40   & \cite{LVD1}\\
MACRO    & 0.26   & 0.054  & 104.545 & 772.727 & 2.78 & 0.5 -- 20   & \cite{MACRO}\\
\end{tabular}}
\end{table} 
For further investigations we have used most recent LVD spectrum presented in Ref.~\cite{LVD1}. The original spectrum of
Ref.~\cite{VZK} (VZK) can not be directly decomposed to~(\ref{Sealds}) because $A_0$ depends logarithmically on energy.
Thus, we were managed to construct a ``simplified'' VZK spectrum with $A_0=0.1258\,k$, where $k$ is the coefficient 
depending on the slant depth $R$, for which underwater flux is calculated, approximately, as:
$k$=0.87 for $R$=(1--7)$\,$km, $k$=0.79 for $R$=(7--11)$\,$km and $k$=0.72 for $R>\,$11 km.

The expression~(\ref{Sealds}) can be rewritten in the shorter form which will be needed for later sections:
\begin{equation}\label{Sealds1}
D(E_{0},\theta)= E_{0}^{-\gamma} \sum_{i={\pi,K}} \frac{D_{0_i}}{1+E_{0}/E^{cr}_{0_i}(\theta)},  
\end{equation}
where $D_{0_\pi}=A_0$ and $D_{0_K}=A_{0}B_0$.

\protect\section{Muon propagation through water}
\label{sec:Mupr}
\protect\subsection{Continuous energy losses}
\label{ssec:CLoss}
For the description of continuous energy losses of muon in water needed for solving the integral equation~(\ref{Ra})
we have done parametrizations based on output data from MUM code.
All formulae for cross sections used in MUM code for muon energy loss computations are described in details in Ref.~\cite{MUM1}.

The energy loss rate per unit of path $x$ is given, conventionally, by the expression
$L(E)=-dE/dx=a(E)+b(E)E$, where $a(E)$ is loss due to ionization and
$b(E)=b_p(E)+b_b(E)+b_n(E)$ is the sum of coefficients for all radiative processes: $e^{+}e^{-}$~pair production ($b_p$),
bremsstrahlung ($b_b$) and photonuclear interaction ($b_n$).   

\begin{figure}[!t]
\center{\mbox{\epsfig{file=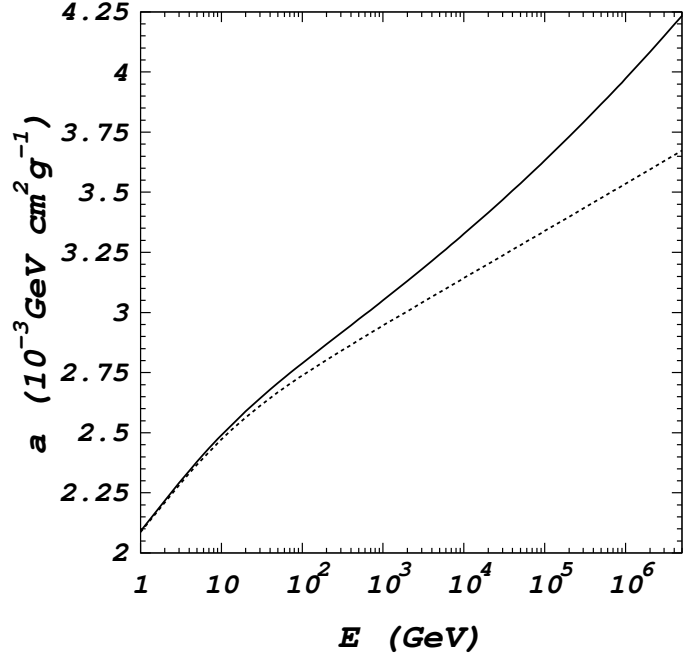,height=9.cm}}}
\protect\caption{
   Ionization muon energy losses in pure water (as MUM code output). Dashed curve -- $a_c$ given by classic
   Bethe-Bloch formula, solid curve -- sum of $a_c$ and $a_e$ (where $a_e$ is loss due to bremsstrahlung with $\gamma$-quantum 
   emission by atomic electron). 
\label{fig:Ionf}}
\end{figure}

Ionization loss was taken as composition of 2 processes (Fig.~\ref{fig:Ionf}),
$a(E)=a_c(E)+a_e(E)$, where $a_c$ is classic ionization calculated using Bethe-Bloch formula
~(Refs.~\cite{lohman,density}) and $a_e$ results from $e$-diagrams for 
bremsstrahlung being treated as a part of ionization process following Refs.~\cite{bremkok,thankok}
($\gamma$-quantum is emitted by atomic electron). 
Taking into account the last process leads to $1.8\,\%,3.4\,\%,5.5\,\%$ increase of ionization loss
for 100 GeV, 1 TeV, 10 TeV muons, correspondingly.
Applying designations used in Ref.~\cite{brembb1} the approximation formula for $a_c$ is given by 
\begin{equation}\label{Ion}
a_c(E)=a_{c_0}+a_{c_1}\ln \left(\frac{W_{max}}{m_{\mu}}\right),~~~~~~~~~  
W_{max}=\frac{E}{1+{m^2_{\mu}}/{(2m_{e}E)}},         
\end{equation}
where $W_{max}$ is maximum energy transferable to the electron and $m_\mu$, $m_e$
are the rest masses of muon and electron. 
The set of cofficients, in units of ($10^{-3}$GeVcm${}^2$g${}^{-1}$), 
\begin{eqnarray*}
            a_{c_0}=2.106,~a_{c_1}=0.0950~ \text{  for } E\leq 45 \text{ GeV,}\\
            a_{c_0}=2.163,~a_{c_1}=0.0853~ \text{  for } E>45\text{ GeV,} 
\end{eqnarray*}
gives the error for parametrization~(\ref{Ion}) smaller than 0.2$\,\%$ for (1--$10^8)\,$GeV range. 
For $a_e$ the following polynomial approximation ($E$ in units of (GeV)): 
\begin{equation}\label{Ionbr}
a_e(E)=3.54 + 3.785\ln E + 1.15\ln^{2}E + 0.0615\ln^{3}E \qquad (10^{-6}\text{GeVcm}^2\text{g}^{-1}),  
\end{equation}
has the error $\lesssim 0.2\,\%$ for (50--$10^8)\,$GeV range.   
Finally, the sum of~(\ref{Ion}) and~(\ref{Ionbr}) has the error of $a(E)$ approximation 
smaller than 0.2$\,\%$ for (1--$10^8)\,$GeV range. 

\begin{figure}[!t]
\center{\mbox{\epsfig{file=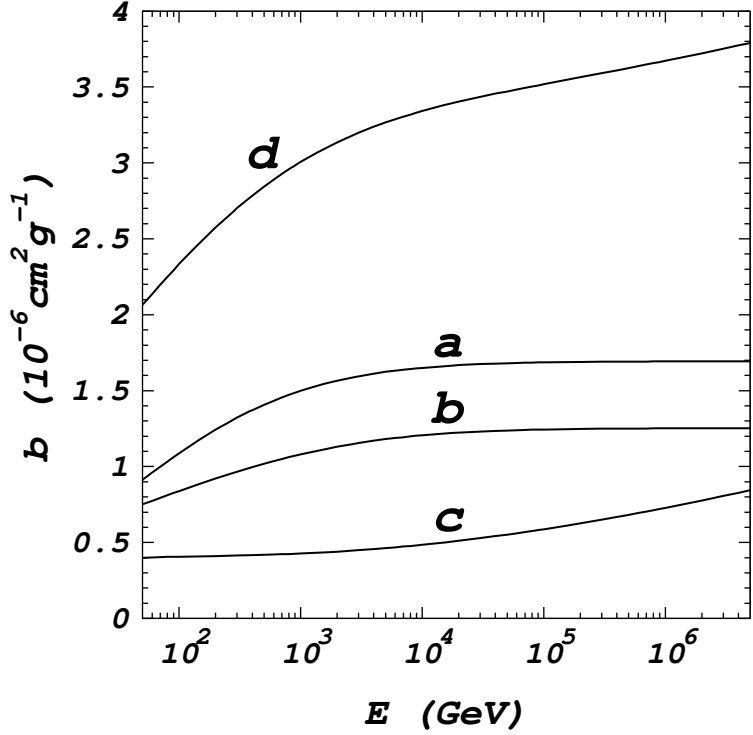,height=10.1cm}}}
\protect\caption{
   The b-terms of muon radiative energy losses in pure water (as MUM code output). Losses due to  
   direct $e^{+}e^{-}$~pair production $b_p$ (curve (a)), bremsstrahlung $b_b$ (b), and photonuclear 
   interaction $b_n$ (c) are shown. Curve (d) is the sum of all b-terms.  
\label{fig:Btermf}}
\end{figure}

Radiative energy losses of muon in water has been calculated using the cross sections from the following works: 
Andreev, Bezrukov and Bugaev (Refs.~\cite{brembb1,brembb2}) for bremsstrahlung, Kokoulin and Petrukhin 
(Refs.~\cite{thankok,pairkok}) for direct $e^{+}e^{-}$~pair production, Bezrukov and Bugaev (Ref.~\cite{phnubb}) 
for photonuclear interaction. Fig.~\ref{fig:Btermf} sketches corresponding b-terms of radiative energy losses. 

In principle MUM code gives values for all b-terms up to $10^9$ GeV.  
Note that the logarithmic rise with an energy of the photoabsorption cross section, used in the model of Ref.~\cite{phnubb}
and resulting in corresponding increase of the $b_n$-term, is experimentally proved already up to the photon energy
$\sim 10^5$ GeV.

For parametrization of b-terms we have used the same functional form as in Ref.~\cite{brembb1} but increased
the power of polynomial to improve accuracy and enlarge the range of application up to 100 PeV without division
on energy subintervals ($E$ in units of (GeV)):
\begin{equation}\label{Bterm}
b_{i}(E)=\sum_{j=0}^4 b_{ij}\ln^{j}E, \qquad \text{where}~i=p,~b,~n.  
\end{equation}
Corresponding coefficients of this decomposition are collected in Tab.~\ref{tab:bttab}.
This fit works with typical errors $\pm$(0.2--0.4)$\%$ within (50--7$\times 10^7$) GeV for all b-terms.\\
\begin{table}[htb]
\protect\caption{ Coefficients $b_{ij}~(10^{-6}\text{cm}^2\text{g}^{-1})$ of the fitting formula~
                 (\protect\ref{Bterm}) for b-terms of muon energy losses in water. Maximum absolute values of relative errors are shown.  
\label{tab:bttab}}
\center{\begin{tabular}{lcrrrrrc}
b-term & subscript i &$b_{i0}~~~~~~$ & $b_{i1}~~~~~~$ & $b_{i2}~~~~~~$ & $b_{i3}~~~~~~$ & $b_{i4}$~~~~~~& Max.err,$\%$ \\\hline
$e^{+}e^{-}$~pairs,~$b_p$ &p&$-11.31\times10^{-1} $ & $7.876 \times10^{-1}$ & $-8.192\times10^{-2}$ & $3.763\times10^{-3}$ & $-6.437\times10^{-5}$ & 0.8\\ 
bremsstrahlung,~$b_b$ &b    &$-1.149\times10^{-1}$ & $2.963 \times10^{-1}$ & $-2.165\times10^{-2}$ & $5.630\times10^{-4}$ & $-2.119\times10^{-6}$ & 0.9\\ 
photonuclear,  ~$b_n$ &n    &$ 3.903\times10^{-1}$ & $ 9.355\times10^{-3}$ & $-3.378\times10^{-3}$ & $4.913\times10^{-4}$ & $-1.216\times10^{-5}$ & 0.6 
\end{tabular}}
\end{table} 
The sum $b(E)$ of fits for b-terms is valid for the energy range (50--$10^8$) GeV with the relative error~$<0.5\,\%$.

\begin{figure}[!t]
\center{\mbox{\epsfig{file=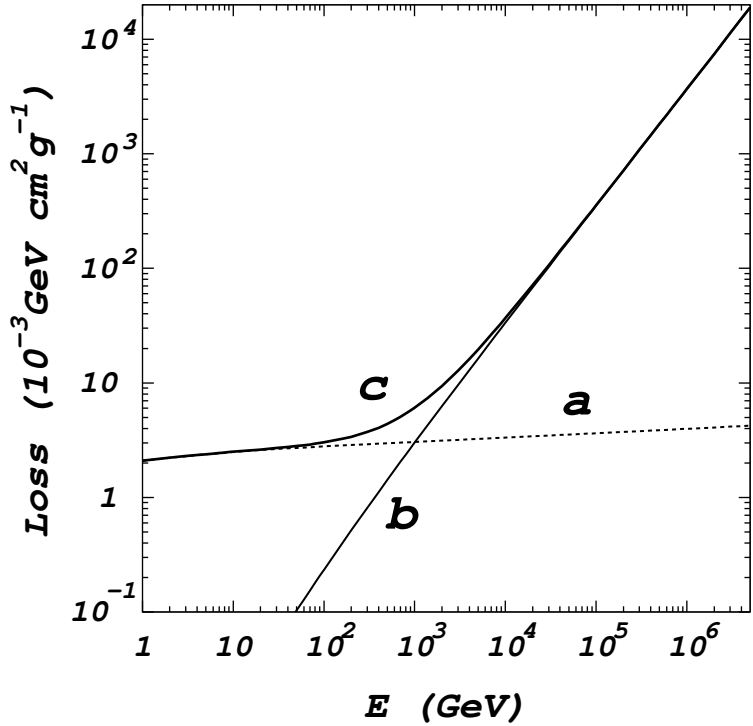,height=10.1cm}}}
\protect\caption{
   Muon energy losses in pure water as a function of energy $E$ (as MUM code output). Curve (a) is loss due to ionization $a(E)$ and 
   (b) is total loss due to all radiative processes $b(E)E$. Curve (c) describes total energy losses $a(E)+b(E)E$.
\label{fig:Totlossf}}
\end{figure}

Fig.~\ref{fig:Totlossf} shows muon energy loss rate in water. Note that the radiative losses account for approximately $5\,\%$ 
of the value of total losses at muon energy 70 GeV, $50\,\%$ at 1 TeV and dominate as $95\,\%$ at 20 TeV.
The sum of all parametrizations~(\ref{Ion}),~(\ref{Ionbr}) and~(\ref{Bterm}) results in the description 
of total energy losses $L(E)$ with excellent accuracy $<0.3\,\%$ with varying sign of error for 
energy range (1--$10^8$) GeV mainly because of b-term error compensations (even for region (1--50) GeV). 
Thanks to this fact just this parametrization of $L(E)$ has been used for the numerical computation of 
integral flux $I_{cl}$ defined by (\ref{Int_cont}) and (\ref{Ra}).

Further, we realized that the total losses $L(E)$ may be described with an accuracy better than $2.5\,\%$ 
(with varying sign of error) for energy range (3--3$\times 10^6$) GeV by 3-slope linear fit  
\begin{equation}\label{Totloss}
            L(E)=\alpha + \beta E, 
\end{equation}
where
$$
\begin{array} {crcrllc}
 \alpha=\alpha_0=& 2.30,&\beta=\beta_0=&15.50&~~~\text { for~~ }&E\leq E_{01}=30.0~~ &\text{GeV,}\\
 \alpha=\alpha_1=& 2.67,&\beta=\beta_1=&3.40 &~~~\text{ for~~ } E_{01} <& E \leq E_{12}=35.3&\text{TeV,}\\
 \alpha=\alpha_2=&-6.50,&\beta=\beta_2=&3.66 &~~~\text{ for~~ } &E > E_{12}=35.3&\text{TeV,}
\end{array}
$$
and $\alpha$ are in units of $(10^{-3}$GeVcm${}^2$g${}^{-1})$ and $\beta$ in $(10^{-6}\text{cm}^2\text{g}^{-1})$.
The energy losses expressed by this parametrizations have sense of effective ones, for example in energy region 30 GeV--35 TeV 
the values $\alpha=2.67$ and $\beta=3.40$ represent effective energy losses due to ionization and radiative processes, correspondingly.
The availability of the linear dependence~(\ref{Totloss}) leads 
to the possibility to derive an analytical formula for underwater integral flux (see Sec.\ref{ssec:Analyt}).   
For simplicity, we will use in the following only 2-slope approximation (with $(\alpha_1,\beta_1)$ and $(\alpha_2,\beta_2)$, 
see the end of Appendix~\ref{app:a2}). 

\protect\subsection{Survival probabilities}
\label{ssec:Sprob}
Probabilities $P(E_0,R,\geq E_f)$, that a muon having starting energy $E_0$,
after getting over the path $R$ will survive with final energy above cut-off $E_f$ (so-called {\it survival probabilities}) 
have been calculated by means of conventional ratio $N(E_0,R,\geq E_f)/N_{tot}(E_0)$ in direct MC-simulations 
by using MUM code with the following inputs.\\

\begin{itemize}
\item[(i)] In addition to fluctuations of all radiative processes (by using the differential cross sections chosen as 
described in preceding subsection) the knock-on electron production was also treated as stochastical (see Ref.~\cite{MUM2}).  
The interactions with fraction of energy lost less than $v_{cut}=$10$^{-3}$ were treated as continuous.   
\item[(ii)] We used $N_{tot}(E_0)$=10$^5$ simulations for each given value $E_0$ of monoenergetic beam resulting 
         in accuracy better than 1$\,\%$ for $P(E_0,R,\geq E_f)\geq\,$0.1.  
         The corresponding MUM-subroutine which watches the muon energy $E_f$ in different points along the path $R$ 
	 in a single simulation act was involved to accelerate considerably the calculations.
\item[(iii)] The grid of simulation parameters $(E_0,R,E_f)$ was sampled to compute underwater angular flux defined by~(\ref{Int_fl})
numerically as: 
\begin{equation}\label{Int_fl1}
I_{fl}(\geq E_f,R,\theta)=\int_{E^{min}_{0}}^{E^{max}_{0}}{P(E_0,R,\geq E_f)D(E_0,\theta)\,dE_0}
\end{equation}
with an systematic error less than 0.5$\,\%$ at the depth of Baikal telescope location $h=$1.15 km for both $E_f$=(1--10$^4$)$\,$GeV
and $\cos\theta$=(0.0495--1.0). It corresponds to slant depths $R=h/\cos\theta$ of (1.15--23.23)$\,$km. The value $E^{min}_{0}$ of minimum
detectable energy was found to be 302 GeV and the value 
$E^{max}_{0}$ needed for avoiding underestimation of integral flux $I_{fl}$ was taken  
as large as 36.3 PeV. Thus, the resulted matrix of survival probabilities
has dimension of $(509 \times 97 \times 81)$ and is computed for a grid of parameters ($E_{0},h\sec\theta,E_{f}$) 
each of them uniformly distributed on: $E_0$=(3.02$\times10^2$--$3.63\times10^{7})\,$GeV with step $\Delta\log_{10} E_0$=0.01,
$\sec\theta$=(1--20.2) with step $\Delta\sec\theta$=0.2, and $E_f$=(1--$10^{4})\,$GeV with step $\Delta\log_{10} E_f$=0.05.
Although the chosen grid was optimized for the vertical depth of $h=$1.15 km,
the resulted survival probabilities may be used for calculations at deeper vertical depths up to slant depths $R$ of 23.23 km.\\ 
\end{itemize}

\begin{figure}[!t]
\center{\mbox{\epsfig{file=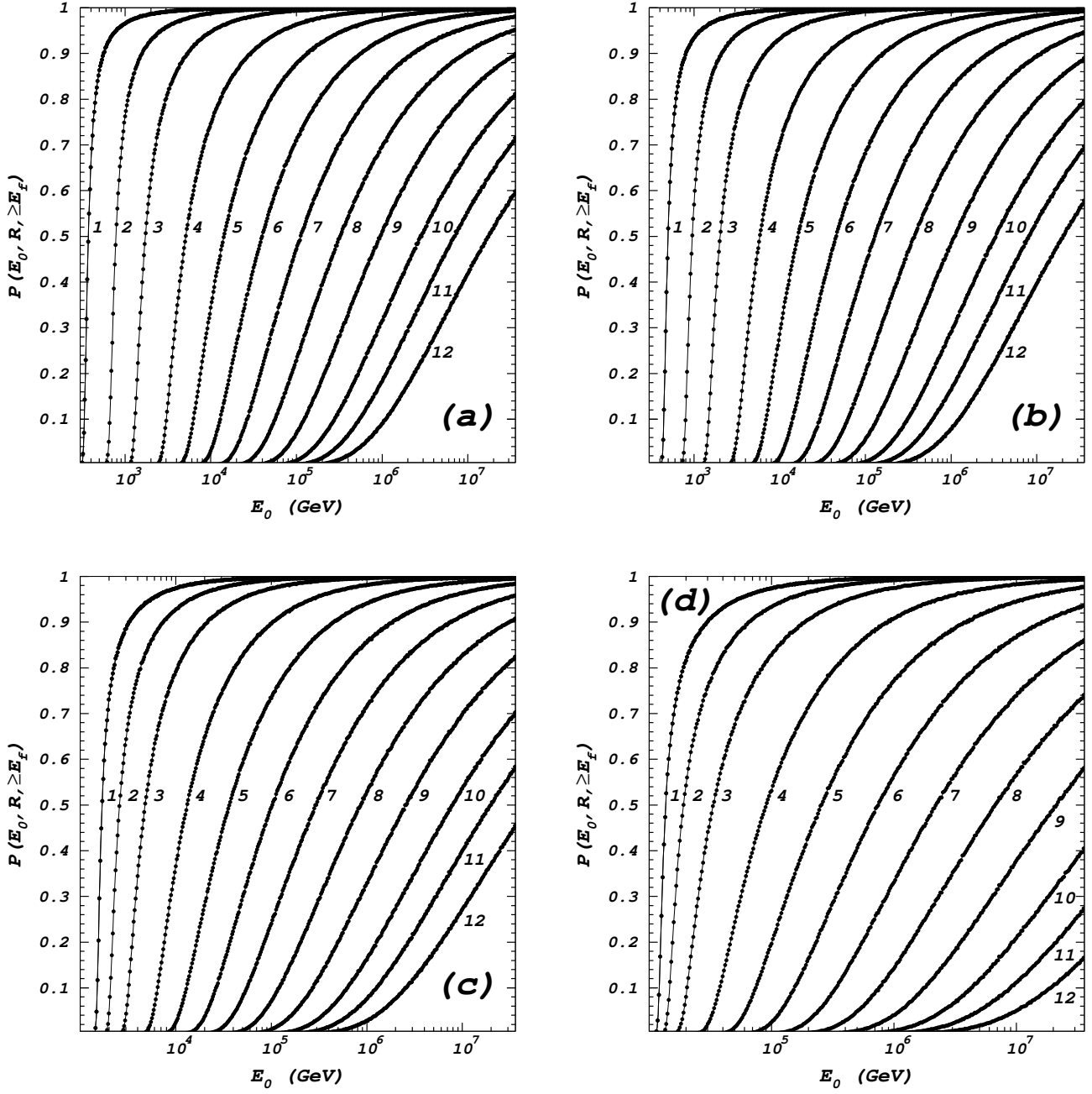,height=18.1cm}}}
\vspace{0.5 cm}
\protect\caption{
Survival probabilities in pure water $P(E_0,R,\geq E_f)$ as a function of starting muon energy $E_0$, resulted  
from MC simulations by using the MUM code as described in Sec.\ref{ssec:Sprob} (dots). 
Solid lines are drawn through the dots by the spline method.
Four pictures are shown
for various cut-off energies $E_f$: 10 GeV (a), 100 GeV (b), 1 TeV (c), and 10 TeV (d), correspondingly. 
Curve labels at each picture correspond to following set of slant depths $R$: 1.15 km (1), 2.07 km (2), 3.45 km (3),
5.75 km (4), 8.05 km (5), 10.35 km (6), 12.65 km (7), 14.95 km (8), 17.25 km (9), 19.55 km (10),
21.39 km (11), and 23.23 km (12), from left to right. 
\label{fig:sp_mumf}}
\end{figure}

\begin{figure}[!t]
\center{\mbox{\epsfig{file=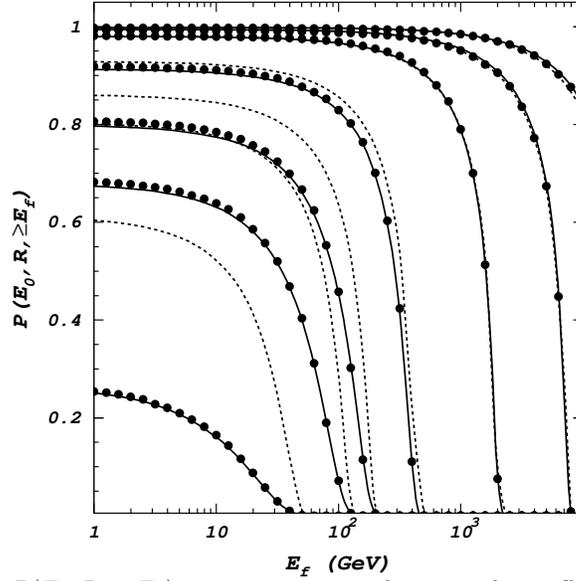,height=8.5cm}}}
\protect\caption{
Survival probabilities $P(E_0,R,\geq E_f)$ in pure water as a function of cut-off muon energy $E_f$ for a given slant depth $R=1.61$ km.
The results of different MC codes of muon propagation are shown. Solid lines -- MUM code, dashed -- PROPMU code,
circles -- MUSIC code. The distributions are given for seven values of starting energy $E_0$ (TeV): 0.5, 0.6, 0.7, 1, 3, 10, and 30, from
left bottom to right top.
\label{fig:sp_allf}}
\end{figure}

The survival probabilities resulted from MUM code are shown in Fig.~\ref{fig:sp_mumf}. 
The sampled comparison of our 
results with those obtained with widely used codes PROPMU~\cite{propmu} and MUSIC~\cite{music} is presented 
in Fig.~\ref{fig:sp_allf}. 
Note, that the greater survival probabilities given by PROPMU for $E_0$=(300--800)$\,$GeV lead to (15--20)$\,\%$ overestimation 
of $I_{fl}(\geq 10~GeV,R,\theta)$ for the slant depths of $R$=(1--2)$\,$km that is critical
for telescopes located at shallow depths (e.g., at lake Baikal). The discrepancy with results given by MUSIC code   
both for survival probabilities and $I_{fl}(\geq E_f,R,\theta)$ is less than 3~$\%$ for a wide range of $E_f$ and $R$.

\begin{figure}[!t]
\center{\mbox{\epsfig{file=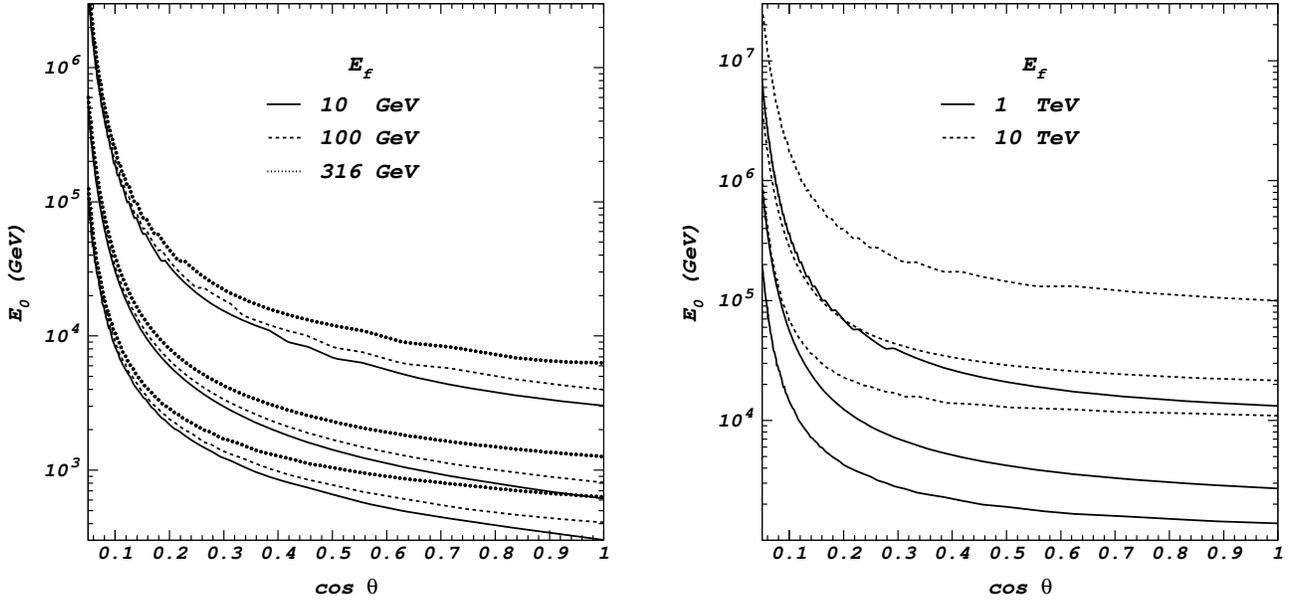,height=9.cm,width=18.cm}}}
\protect\caption{
The values of starting muon energy $E_0$ at a sea level which result in passing till underwater vertical depth of 1.15 km 
at various zenith angles and cut-off energies $E_f$. Results obtained with MUM code with taking into account loss fluctuations are given.
Solid curves are shown for $E_f$ of 10 GeV (left picture) and 1 TeV (right picture). 
Dashed curves are shown for $E_f$ of 100 GeV (left picture) and 10 TeV (right picture). 
Dotted  curves are shown for $E_f$ of 316 GeV at left picture. 
For each $E_f$ we present the set of three curves corresponding to following meaning of $E_0$: $E^{min}_{0}$, 
$\overline E_{0}$ and $E^{max}_{0}$ from bottom to top. 
$E^{min}_{0}$ is a minimal detectable starting energy $E_0$ given by survival probability of $10^{-4}$. 
$\overline E_{0}$ is a mean sea level energy
calculated using NSS spectrum. $E^{max}_{0}$ is that value of upper limit of the integral~(\protect\ref{Int_fl1}), which ensures
99.5$\,\%$ of underwater flux, calculated using NSS sea level spectrum with $E^{max}_{0}$=36.3 PeV.     
\label{fig:Estartf}}
\end{figure}

Fig.~\ref{fig:Estartf} sketches the values of starting muon energy $E_0$ at a sea level 
which result in passing till underwater vertical depth of 1.15 km 
at various zenith angles and cut-off energies $E_f$. 
\protect\section{Underwater spectra of atmospheric muons}
\label{sec:Underw}
\protect\subsection{Correction factor}
\label{ssec:Numcalc}
The influence of fluctuations of muon losses in matter (mainly due to radiative processes) results in
that the real integral flux $I_{fl}$ is generally greater than $I_{cl}$ calculated in the approximation of continuous losses. 
In this work we propose to allow for the influence of energy loss fluctuations on the value of
angular flux in matter by means of correction factor expressed by the ratio   
\begin{equation}\label{Cf1}
C_{f}(\geq E_f,R,\theta)=\frac {I_{cl}(\geq E_f,R,\theta)} {I_{fl}(\geq E_f,R,\theta)}.  
\end{equation}
Here $I_{fl}$ is given by~(\ref{Int_fl}) and $I_{cl}$ is given by~(\ref{Int_cont}). 
In the assumption when the same differential cross sections are used for computing both the numerator and denominator
of ratio~(\ref{Cf1}), one may expect that this factor depends only weakly on sea level angular spectrum.  
For numerical calculations of the correction factor we used total continuous energy 
losses defined by sum of~(\ref{Ion}),~(\ref{Ionbr}) and~(\ref{Bterm}), and survival probabilities
described in preceding section.  

\begin{figure}[!t]
\centering
  \mbox{\epsfig{file=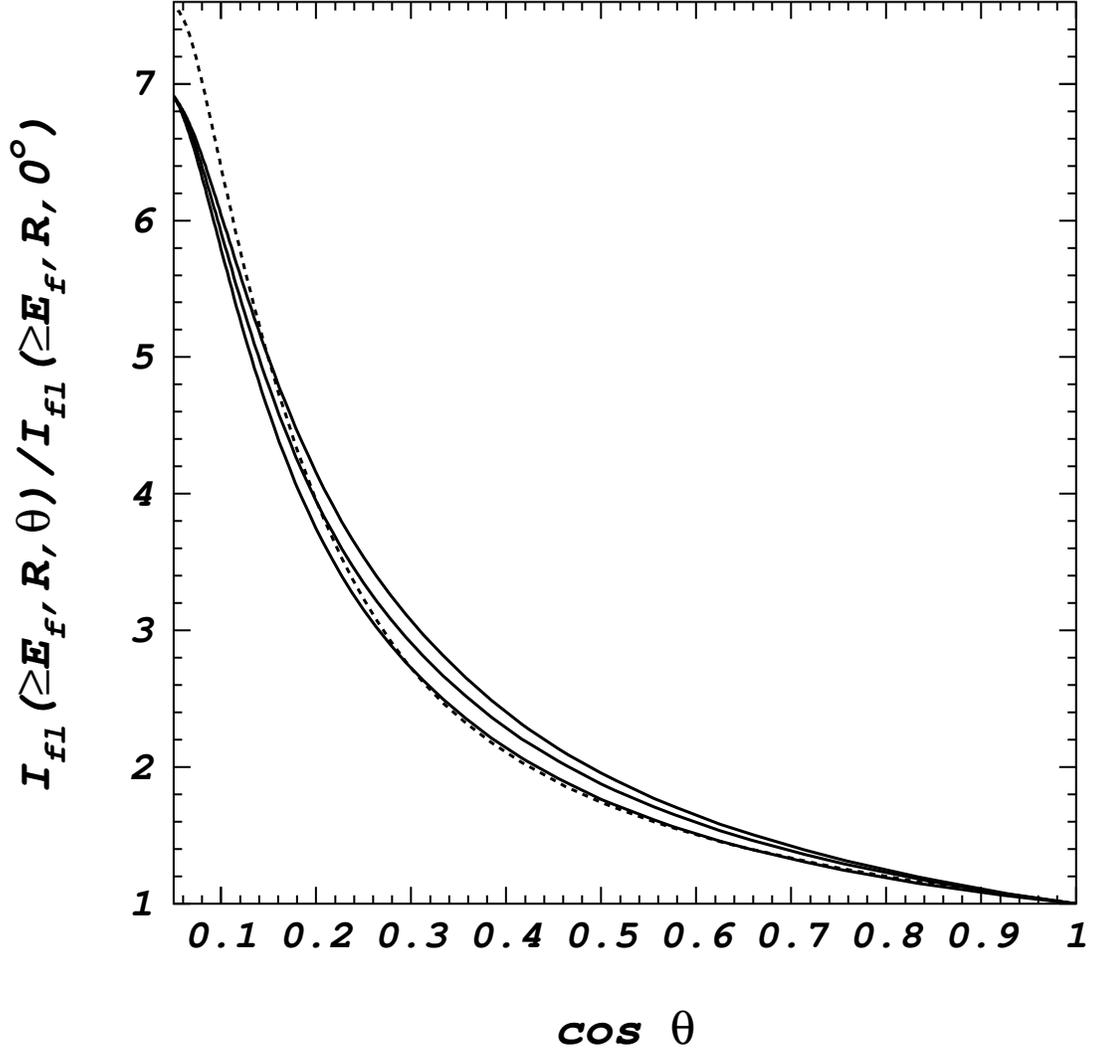,height=15.cm}}
\protect\caption{
The ratio of underwater integral flux expected at vertical depth $h$=1.15 km in water at various zenith angles to
that expected for vertical direction for the corresponding slant depths $R=h/\cos\theta$.
Solid curves result from numerical computations allowing for loss fluctuations by using of NSS sea level spectrum.
They are given for three values of cut-off energy $E_f$: 10 GeV, 1 TeV and 10 TeV, from bottom to top.  
Dashed curve is derived from sea level spectrum of Ref.~\protect\cite{VZK} with an angular distribution from Ref.~\protect\cite{Volkova75}
and is shown for $E_f$=10 GeV.
\label{fig:otn}}
\end{figure}

\begin{figure}[!t]
\center{\mbox{\epsfig{file=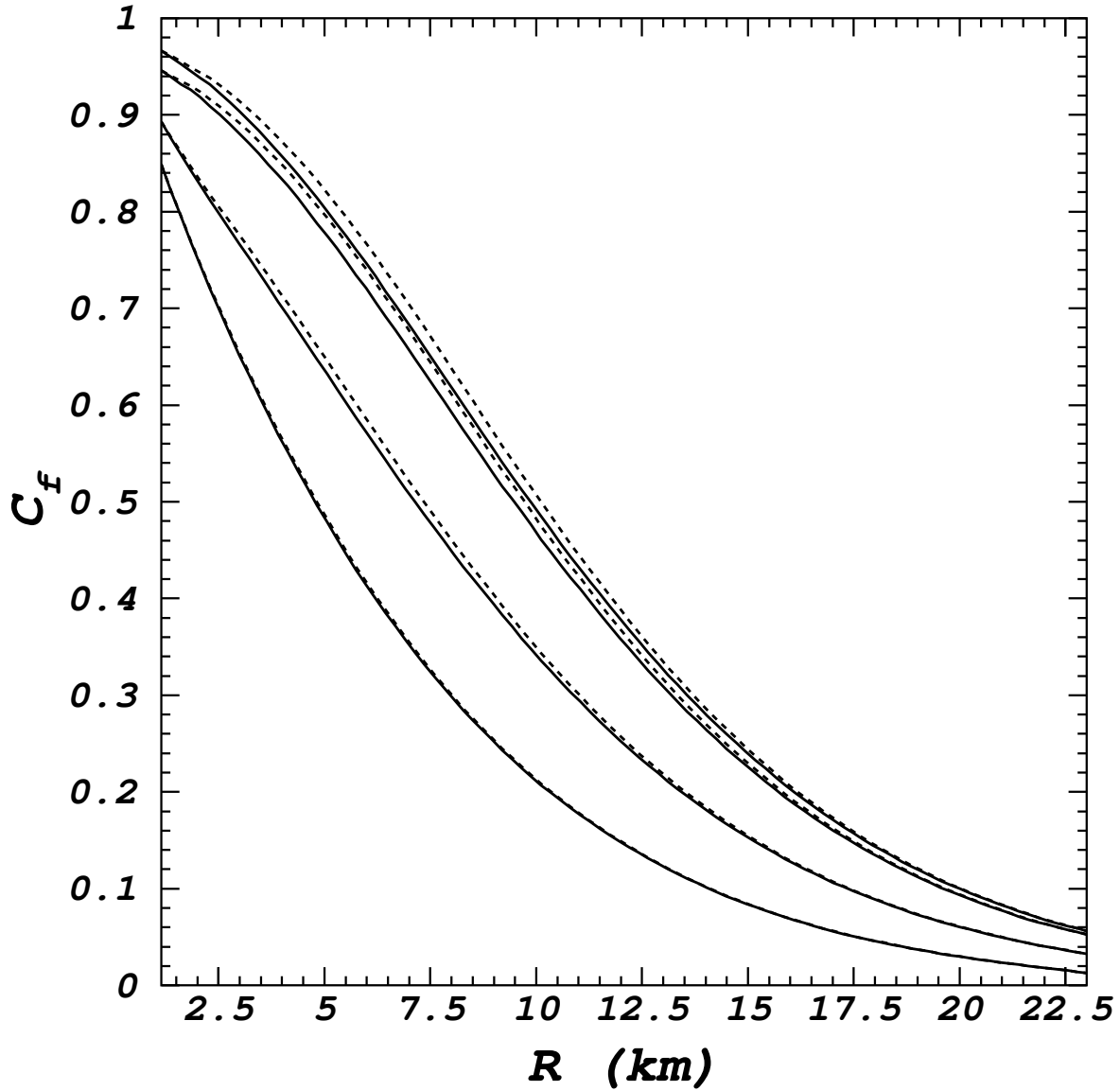,height=17.cm}}}
\protect\caption{ 
Correction factor $C_f$ as a function of slant depth $R$ in pure water. 
The results obtained using sea level spectrum defined by Eq.~(\protect\ref{BKfit}) are given.
Solid curves correspond to numerical calculations for vertical case $\theta=0^\circ$. 
Dashed curves describe the correction factor computed at vertical depth $h$ of 1.15 km for various zenith angles
as a function of slant depth defined by $R=h/\cos\theta$. 
Both solid and dashed curves are shown for four values of cut-off energy 
$E_f$: 10 GeV, 100 GeV, 1 TeV, and 10 TeV, from top to bottom.
\label{fig:cf_bk}}
\end{figure}

The value of underwater integral flux for a given slant depth $R$ at zenith angle $\theta$ is greater than the flux calculated for 
the same $R$ for vertical direction as is shown in Fig.~\ref{fig:otn}. 
It results from angular distribution of sea level spectrum shown in Fig.~\ref{fig:Ang}. At the same time the values of correction
factors calculated for the same slant depth $R$ at vertical direction and at zenith angle $\theta$ differ weakly. 
It is illustrated in Fig.~\ref{fig:cf_bk}, where one can see that $C_{f}(\geq E_f,R,0^{\circ})$ differs from  
$C_{f}(\geq E_f,R,\arccos h/R)$ maximum on 3.3$\,\%$ for $E_f>$10 GeV at vertical depth $h$ of 1.15 km. 
It appears that with acceptable accuracy the correction factor depends on slant depth $R$ only, rather than on $R$ and $\theta$
separately. 

The dependencies of correction factor on $E_f$ and $R$, calculated for sea level spectrum
~(\ref{BKfit}), represent the set of rather smooth curves (shown in Fig.~\ref{fig:cf_bk}) 
and it is possible to approximate this factor with formula  
\begin{equation}\label{CF}
C_f(\geq E_f,R,\theta)=\sum_{i=0}^4 ( \sum_{j=0}^4 c_{ij}\log^{j}_{10}\,E_{f} )R^i. 
\end{equation}
Here cut-off energy $E_f$ is expressed in (GeV) and slant depth $R$ is in (km) with the coefficients $c_{ij}$
collected in Table~\ref{tab:cftab}. When using~(\ref{CF}) for cut-off energies $E_f<$10 GeV one should substitute value of $E_f$=10 GeV.

Formula~(\ref{CF}) can be applied for any geometrical shape of the surface. Right hand side of~(\ref{CF}) depends on
$\theta$ because, generally, $R=R(\theta)$. So, in the particular case of a flat surface the angular dependence of the
correction factor appears, in our approximation, only through the relation  
$$R=\frac {h}{\cos\theta},$$
where $h$ is a vertical depth. 

The accuracy of formula~(\ref{CF}) for $E_f$=(1--100)$\,$GeV is better than $\pm2\,\%$ for slant depths $R$ as large as 22 km and
is not worse than $\pm3\,\%$ for $E_f$=1 TeV up to $R$=17 km and for $E_f$=10 TeV up to $R$=15 km.
\begin{table}[htb]
\protect\caption{ Coefficients $c_{ij}$ of the fitting formula~(\protect\ref{CF}) for correction factor calculated for vertical
                  basic sea level spectrum~(\protect\ref{BKfit}) used in this work.    
\label{tab:cftab}}
\center{\begin{tabular}{crrrrr}
subscript $i$  & $c_{i0}$~~~~~~~ & $c_{i1}$~~~~~~~ & $c_{i2}$~~~~~~~ & $c_{i3}$~~~~~~~ & $c_{i4}$~~~~~~~ \\\hline
0              &  $ 6.3045 \times10^{-1}$ & $ 6.6658 \times10^{-1}$ & $-4.5138 \times10^{-1}$ & $ 1.2441 \times10^{-1}$ & $-1.1904 \times10^{-2}$ \\ 
1              &  $ 2.0152 \times10^{-1}$ & $-4.2990 \times10^{-1}$ & $ 3.2532 \times10^{-1}$ & $-1.0265 \times10^{-1}$ & $ 1.0751 \times10^{-2}$ \\ 
2              &  $-3.3419 \times10^{-2}$ & $ 5.1833 \times10^{-2}$ & $-3.9229 \times10^{-2}$ & $ 1.2360 \times10^{-2}$ & $-1.2911 \times10^{-3}$ \\ 
3              &  $ 1.6365 \times10^{-3}$ & $-2.3645 \times10^{-3}$ & $ 1.7775 \times10^{-3}$ & $-5.5495 \times10^{-4}$ & $ 5.7557 \times10^{-5}$ \\ 
4              &  $-2.6630 \times10^{-5}$ & $ 3.7770 \times10^{-5}$ & $-2.8207 \times10^{-5}$ & $ 8.7275 \times10^{-6}$ & $-8.9919 \times10^{-7}$ \\ 
\end{tabular}}
\end{table} 
Fig.~\ref{fig:cf_bk} shows that for $E_f<\,$100 GeV the total energy loss may be treated as quasi-continuous  
(at level of $C_f>\,$0.9) only for slant depths $R<\,$2.5 km but for $E_f$=10 TeV the fluctuations should be taken into account
at level of 15$\,\%$ already for slant depth as small as $R$=1 km. 
For slant depths larger than 10 km $I_{fl}$ and $I_{cl}$ differ more than on a factor of 2.

\begin{figure}[!t]
\centering
  \mbox{\epsfig{file=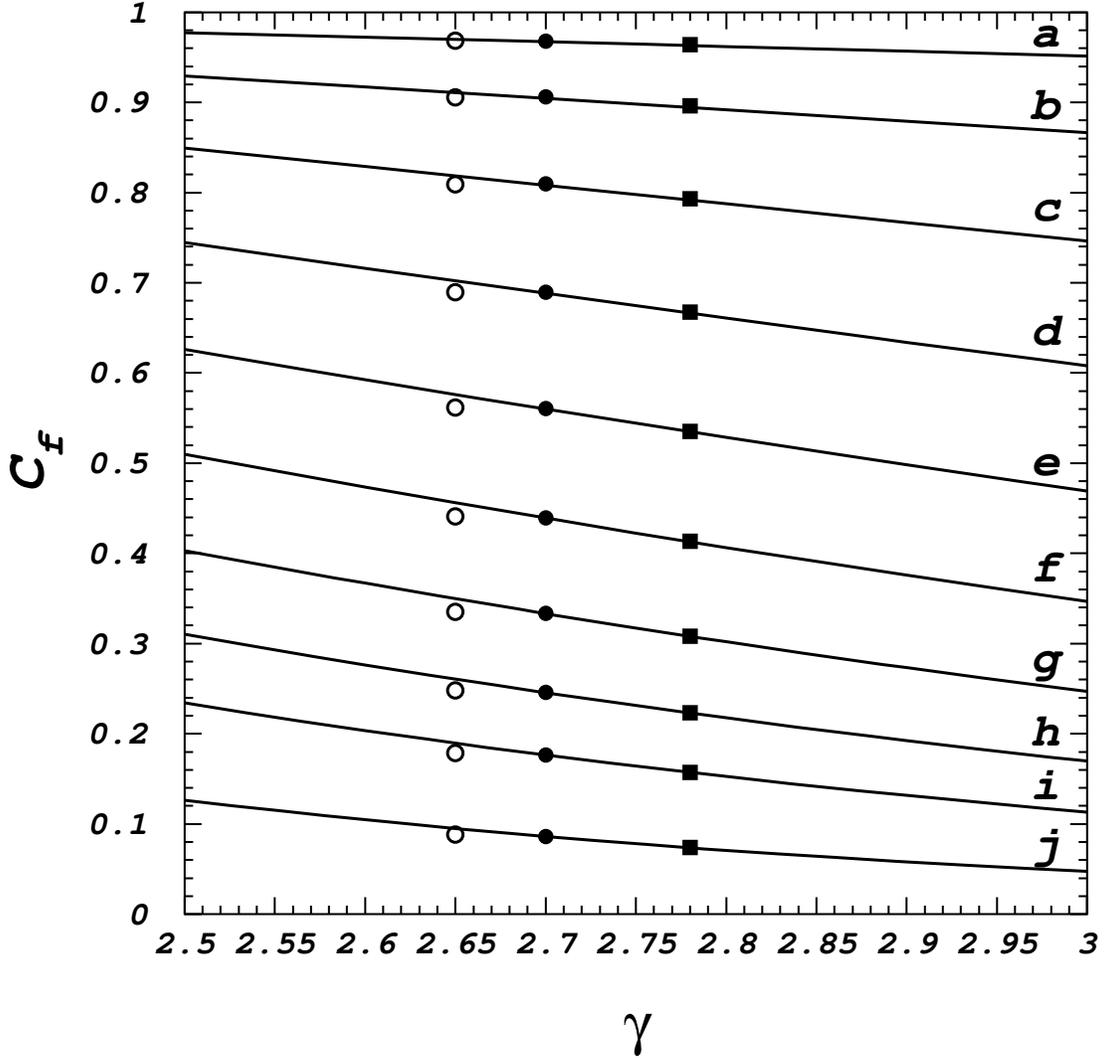,height=15.cm}}
\protect\caption{ 
Correction factor $C_f$ as a function of spectral index $\gamma$ of sea level spectrum for various depths in water for vertical direction.
The distributions for cut-off energy $E_f$=10 GeV are given.
Solid curves correspond to numerical computations by using sea level spectrum defined by Eq.~(\protect\ref{BKfit}) with varying 
spectral index $\gamma$. Open circles correspond to numerical computations using ``simplified'' VZK sea level spectrum,
closed circles -- Gaisser's sea level spectrum, squares -- MACRO sea level spectrum (see Table~\protect\ref{tab:seal_all}). 
All distibutions are shown for the following values of vertical depth in pure water: 1.15 km (a), 3 km (b), 5 km (c), 7 km (d),
9 km (e), 11 km (f), 13 km (g), 15 km (h), 17 km (i), and 21 km (j), from top to bottom.
\label{fig:cf_gamma}}
\end{figure}

The dependence of correcton factor $C_f$ on different sea-level vertical spectra from the Table~\ref{tab:seal_all} 
is illustrated by Fig.~\ref{fig:cf_gamma}. The correction factors calculated for $E_f$=10 GeV using sea level spectrum   
~(\ref{BKfit}) with spectral index $\gamma$ of 2.5 and 3.0 (instead of 2.72) differ more than on a factor of 2 starting from 
slant depth of $R$=12 km. Nevertheless, the values of $C_f$ calculated using sea level spectra having $\gamma$=(2.65--2.78) 
are already within $\pm5\,\%$ corridor. For $E_f$=1 TeV this corridor is larger on $2\,\%$.
This fact results in the possibility to extrapolate the parametrization~(\ref{CF}) based on sea level spectrum having $\gamma$=2.72 
to other spectra from Table~\ref{tab:seal_all} at least up to slant depths of (12--13) km without introduction 
of additional spectral corrections.  
\protect\subsection{Comparison of numerical and analytical calculations}
\label{ssec:Analyt}
Taking into account the results of Appendix~\ref{app:a2} and preceding subsection we present finally the analytical expressions
for calculations of underwater angular flux above cut-off energy $E_f$ for a slant depth $R=h/ \cos \theta$ seen at vertical depth $h$ 
at zenith angle $\theta$ and allowing for the fluctuations of energy loss:
\begin{equation}\label{ad1}
F_{fl}(\geq E_f,R,\theta)=\frac {F_{cl}(\geq E_f,R,\theta)} {C_{f}(\geq E_f,R,\theta)},  
\end{equation}
where correction factor is expressed by~(\ref{CF}) and angular flux $F_{cl}(\geq E_f,R,\theta)$ based on effective 
linear continuous energy losses $\alpha + \beta E$ having 2 slopes, is calculated by the following rule: 
\begin{equation} \label{ad2}
F_{cl}(\geq E_f,R,\theta) = \left\{
\begin{aligned} F_{cl}(\geq E_f,R,\theta;~\alpha_{1},\beta_{1})
                         & \text{~~for~}  R \leq R_{12}, \\
                F_{cl}(\geq E_{12},(R-R_{12}),\theta;~\alpha_{2},\beta_{2})
                         & \text{~~for~}  R > R_{12}. 
\end{aligned}\right.
\end{equation}
Here $E_{12}$ is the energy in the point of slope change from $(\alpha_1,\beta_1)$ to $(\alpha_2,\beta_2)$ and 
$R_{12}$ is the muon path from the energy $E_{12}$ till $E_f$ which is given by
$$R_{12}=\frac {1}{\beta_1}\,\ln \biggl( \frac {\alpha_{1}+E_{12}\beta_{1}}{\alpha_{1}+E_{f}\beta_1} \biggr).$$ 

As is shown in Appendix~\ref{app:a2} the formula for integral muon angular flux in the assumption of linear 
continuous energy losses is as follows:
\begin{equation}\label{ad3}
F_{cl}(\geq E_f,R,\theta;~\alpha,\beta)=\frac{e^{-\beta R\gamma}}{\gamma}\sum_{i={\pi,K}} D_{0_i} E^{cr}_{0_i}(\theta) 
(E_f+y_i)^{-\gamma} (1-z_i)^{1-\gamma}\,\text{S}(z_{i},\gamma), 
\end{equation}
where subscript $i$ stands over both pion ($\pi$) and kaon ($K$) terms and 
\begin{eqnarray*}
y_i = \frac{\alpha}{\beta}\,(1-e^{-\beta R}) + E^{cr}_{0_i}(\theta)\,e^{-\beta R},
\qquad z_i = \frac {E^{cr}_{0_i}(\theta)\,e^{-\beta R}}{ E_{f}+y_i },
\qquad E^{cr}_{0_i}(\theta)=\frac {E^{cr}_{0_i}(0^\circ)}{\cos\theta^*},\\
\text{S}(z,\gamma) = 1+\frac{z}{\gamma+1}+\frac{2z^2}{(\gamma+1)(\gamma+2)}+ 
\frac{6z^3}{(\gamma+1)(\gamma+2)(\gamma+3)} + \dots~. 
\end{eqnarray*}
The 5 parameters ($D_{0_\pi},D_{0_K},E^{cr}_{0_\pi}(0^\circ),E^{cr}_{0_K}(0^\circ),\gamma$) are those of the sea level spectrum~(\ref{Sealds})
taking into account the notations in~(\ref{Sealds1}):
\[
D_{0_\pi}=A_0, \qquad D_{0_K}=A_{0}B_0. 
\]

The corresponding angular distrubution should be introduced using an analytical description of effective cosine $\cos\theta^*$  
taking into account the sphericity of atmosphere. It should be noted that the description of underwater angular flux with the
5 parameters of a sea level spectrum gives the possibility of their direct best fit by using the 
experimental underwater distribution.

The flux value in~(\ref{ad3}) is expressed in units of (cm$^{-2}$s$^{-1}$sr$^{-1}$) and all energies are in (GeV),
slant depth $R$ in units of (g$\,$cm$^{-2}$), loss terms $\alpha$ and $\beta$ are in units 
of $(10^{-3}$GeVcm${}^2$g${}^{-1})$ 
and $(10^{-6}\text{cm}^2\text{g}^{-1})$, correspondingly. Note that when calculating correction factor with~(\ref{CF}) 
range $R$ is presented in units of (km).

To simplify the formula we have used only two slopes from approximation~(\ref{Totloss}) when substituting in~(\ref{ad2}):  
($\alpha_1$=2.67, $\beta_1$=3.40) and ($\alpha_2$=$-$6.5, $\beta_2$=3.66) with $E_{12}$=35.3 TeV. To examine the angular behaviour of 
a flux given by the formula~(\ref{ad1})
by means of the comparison with numerical calculations based on NSS spectrum, we have used (according to the fit~(\ref{BKfit})),
 instead of Eq.~(\ref{cren}), the modified expression for the critical energies 
$E^{cr}_{0_{\pi,K}}(\theta)=E^{cr}_{0_{\pi,K}}(0^\circ)/\cos\theta^{**}$, where $\cos\theta^{**}$
is given by the expression~(\ref{Efcos}). 
The dependence on the sea level spectrum choice was investigated for vertical direction only, by using the
sea level spectra collected in Table~\ref{tab:seal_all}. 

The examination of~(\ref{ad3}) showed rather quick convergence of series S$(z,\gamma)$ with increase of $R$ and $E_f$. 
Therefore, for the accuracy of $F_{cl}$ computation better than 0.1~$\%$ it is quite enough to take only 
four first terms of this series (up to $z^3$) for all values $R>$~1 km and $E_f$ in (1--10$^{4}$) GeV. Even using the two
terms leads to the accuracy of 1.3~$\%$ for ($R$=1.15 km, $E_f$=1 GeV) and $<$0.5$\,\%$ for ($R>\,$2.5 km, $E_f>\,$1 GeV).

\begin{figure}[!t]
\center{\mbox{\epsfig{file=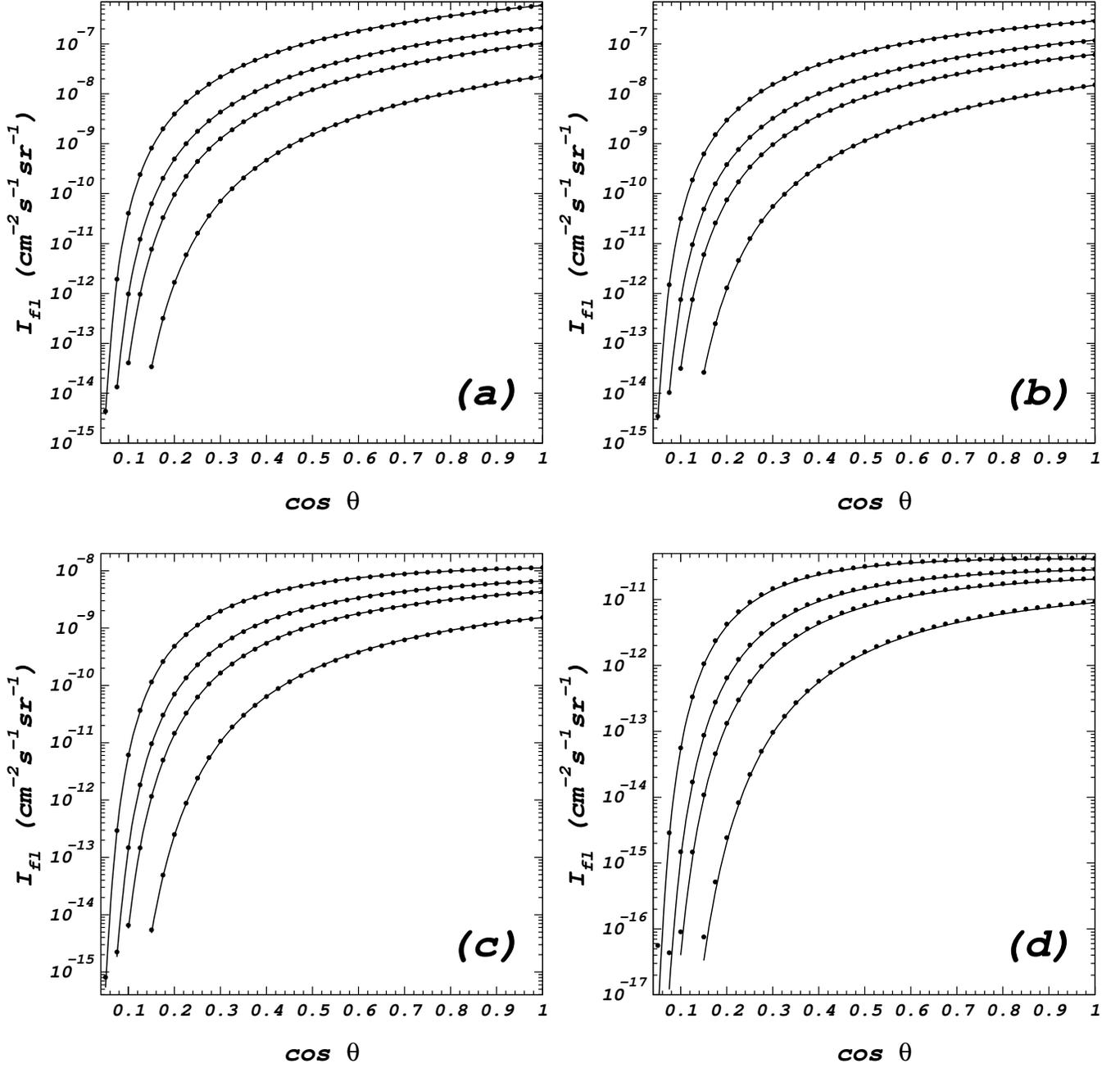,height=18.cm}}}
\protect\caption{ 
Underwater integral muon flux allowing for loss fluctuations as a function of zenith angle at different vertical depths.
Four pictures are shown for various cut-off energies $E_f$: 10 GeV (a), 100 GeV (b), 1 TeV (c), and 10 TeV (d), correspondingly. 
Four curves at each picture correspond to vertical depths $h$: 1.15 km, 1.61 km, 2.0 km, and 3.0 km, from top to bottom. 
Solid curves result from numerical computations by using the NSS sea level spectrum and dotted ones from analytical
expression~(\protect\ref{ad1}) by using the sea level spectrum~(\protect\ref{BKfit}). 
\label{fig:Ifl_all}}
\end{figure}

\begin{figure}[!t]
\centering
  \mbox{\epsfig{file=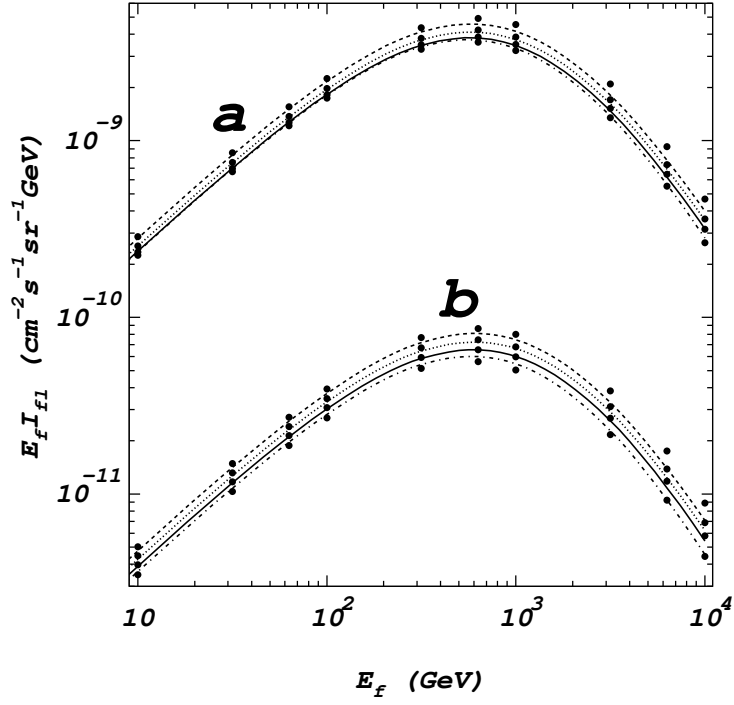,height=9.7cm}}
\caption[1]{\small 
Underwater vertical integral muon flux allowing for loss fluctuations as a function of cut-off energy $E_f$ for
various sea level spectra presented in Table~\protect\ref{tab:seal_all}. Two set of curves are given for two vertical depths $R$: 10 km (a)
and 15 km (b).  
Dashed curves result from numerical computations by using of VZK sea level spectrum~\protect\cite{VZK},
dotted -- Gaisser's sea level spectrum, solid -- parametrization~(\protect\ref{BKfit}) of present work, 
dash-dotted -- LVD sea level spectrum.  
Closed circles result from analytical expression~(\protect\ref{ad1}) with using the same sea level spectra.
\label{fig:Fvert}}
\end{figure}

Fig.~\ref{fig:Ifl_all} shows the comparison of underwater angular integral fluxes allowing for loss fluctuations 
at different basic depths $h$ (of location of existing and planned telescopes) calculated both numerically~(\ref{Int_fl1}) for NSS spectrum 
and analytically~(\ref{ad1}) for the spectrum given by~(\ref{BKfit}). Note the coincidence of our numerical result obtained using 
Monte Carlo survival probabilities with that calculated using analytical method in the first paper of Ref.~\cite{Sineg} at level of $\pm5\,\%$ for basic
depth $h$ of 1.15 and 1.61 km for $E_f$=(10--20) GeV up to $\cos\theta=$(0.1--0.15).  
Fig.~\ref{fig:Fvert} gives the comparison of numerical results and that given by the formula for different vertical sea level spectra
for 2 selected depths of interest: 10 km and 15 km. 

We realized that the error given by formula~(\ref{ad1}) for     
all mentioned sea level spectra is within the corridor of $\pm$(4--6)$\,\%$ for all cut-off 
energies $E_f$=(1--10$^{3})\,$GeV and
slant depths $R$=(1--16)$\,$km (corresponding angle is expressed by $\cos\theta=h/R$ for a given vertical depth $h$). This is proved
for $h$ in a range (1--3)$\,$km. 
For bigger cut-offs of $E_f$=(1--10)$\,$TeV the corridor of errors is $\pm$(5--7)$\,\%$ for $R$=(1--13)$\,$km. Note that for the
sea level spectrum~(\ref{BKfit}), just used for $C_f$ parametrization, the errors are smaller on 2$\,\%$. 

Note that the expression~(\ref{ad1}) may be directly used for an ice after substitution $R \to R/ \rho$, with $\rho$ being the ice density
and with an additional error of $\sim2\,\%$ for a sea water. In spite of seeming complexity of the formulae~(\ref{ad1}),
~(\ref{ad2}) and~(\ref{ad3}) they may be easily programmed.

\begin{figure}[!t]
\centering
  \mbox{\epsfig{file=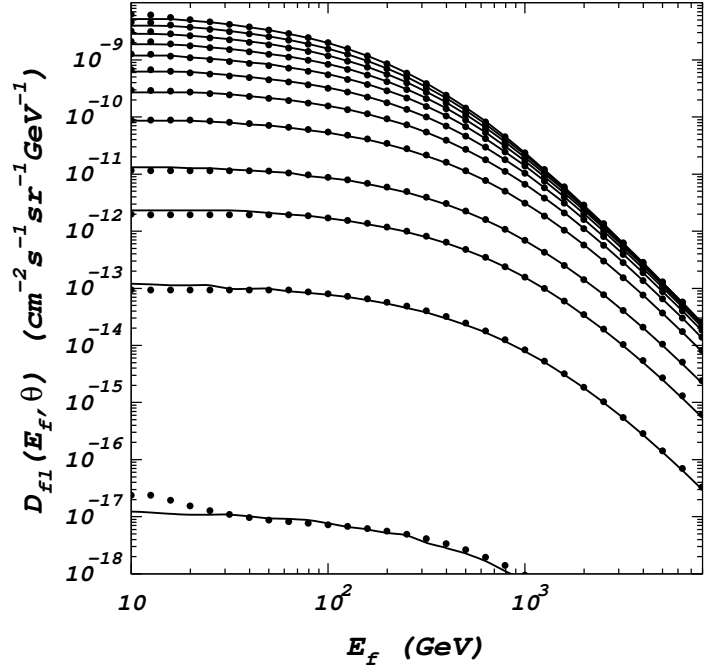,height=9.5cm}}
\protect\caption{ 
Underwater differential muon spectrum allowing for loss fluctuations as a function of energy $E_f$ at vertical depth of 1.15 km.
The curves are given for twelve zenith angles $\cos\theta$: 1.0, 0.9, 0.8, 0.7, 0.6, 0.5, 0.4, 0.3, 0.2, 0.15, 0.1, and 0.05, 
from top to bottom.  
Solid curves result from numerical differentiation of integral flux expressed by~(\protect\ref{Int_fl1}) and based on NSS 
sea level spectrum. Dotted curves result from numerical differentiation of the analytical expression~(\protect\ref{ad1}) 
based on the sea level spectrum~(\protect\ref{BKfit}). 
\label{fig:diff_ef}}
\end{figure}

The validity of proposed formula up to cut-off energies 10 TeV allows a calculation of underwater angular differential
spectrum $D(E_{f},R,\theta)$ by means of numerical differentiation of expression~(\ref{ad1}). It leads to rather appropriate 
results up to slant depths (11--12)$\,$km. 
We illustrate this in Fig.~\ref{fig:diff_ef} by comparison the underwater angular spectra calculated by numerical 
differentiation of integral fluxes given both by~(\ref{Int_fl1}) for NSS sea level spectrum 
and analytically~(\ref{ad1}) for spectrum given by~(\ref{BKfit}). For a vertical depth $h$=1.15 km it yields to errors
$\pm4\,\%$ for $E_f$=(20--$8\times10^{3})\,$GeV for the angles corresponding to slant depths $R=h/\cos\theta$ of (1--3)$\,$km
and $\pm$(6--8)$\,\%$ for $E_f$=(30--$5\times10^{3})\,$GeV for the slant depths (3--12)$\,$km. Even for $R$=23.2 km the result
is still valid within $\pm10\,\%$ but for the very narrow energy region $E_f$=(90--300)$\,$GeV. 
\protect\section{Conclusions}
\label{sec:Concl}
The analytical expression presented in this work allows to estimate for fluctuating losses the integral flux of
atmospheric muons in pure water expected for different zenith angles, $\cos\theta$=(0.05--1.0), at various vertical
depths at least of $h$=(1--3)$\,$km for different parametrizations of the sea level muon spectra. 
The errors of this expression are estimated to
be smaller than $\pm$(4--6)$\,\%$ for cut-off energies ranged in $E_f$=(1--10$^{3})\,$GeV and slant depths 
in $h/\cos\theta$=(1--16)$\,$km. The main advantage of the presented formula consists in the possibility of the direct best fit of 5
parameters of parent sea level spectrum using angular distribution of underwater integral flux measured experimentally at a given
vertical depth.  

The validity of this analytical expression with an accuracy of $\pm$(5--7)$\,\%$ for $E_f$=(10$^3$--10$^4$)$\,$GeV 
and slant depths of (1--12)$\,$km gives also the possibility of estimation the angular underwater differential 
spectrum (by means of numerical differentiation) with error smaller than $\pm$(6--8)$\,\%$ for 
energies of (30--5$\times10^{3}$)$\,$GeV.

The accuracy of the presented parametrization of the correction factor as a function of $E_f$ and slant depth $R$   
is rather high and is about $\pm5\,\%$ for all angles and kinds of the sea level spectrum (assuming that the spectral
index $\gamma$ is approximately within (2.65--2.78)). It results in the possibility to use it for an estimating
numerically from various sea level spectra the value of an angular integral flux allowing for fluctuations of losses without
direct Monte Carlo simulations.

The results of this work may be used directly to the estimations in ice (substituting a slant depth in ice in units of water equivalent)
and with additional error $\sim 2\,\%$ for a sea water. The proposed method may be adapted to estimations in rock after
corresponding description of the correction factor and continuous effective losses.

\acknowledgments
We are grateful to V. A. Naumov for useful advices and to S. I. Sinegovsky and T. S. Sinegovskaya for making available
the muon sea level spectrum data tables. We acknowledge useful discussions with A. V. Butkevich.
\protect\appendix
\section{Geometrical effective cosine due to spherical atmosphere}
\label{app:a1}
As was shown in Ref~\cite{BN} for the spherical isotermal atmosphere the effective cosine is defined by:
\begin{equation}\label{Efcos1}
\cos\theta^{*}=\frac {1}{\exp (\xi^2)(1-\text{erf}(\xi))} \sqrt{\frac {2H_0}{\pi(R_{E}+H)} },
\end{equation}
where
\begin{equation}\label{Efcos2}
\xi^2=\frac{1}{1+H/R_E}\Biggl( \frac{\cos^2{\theta}R_{E}}{2H_0}+\frac{H}{H_0}\biggl(1+\frac{1}{2R_E}\biggr)\Biggr).       
\end{equation}
$R_E$ is the Earth radius (we used mean value $6367.554\,$km), $H_0=6.4385\,$km for T$=220^\circ\,$K, erf is the standard integral of probability.    
$H$ has sense of the effective height of muon generation in spherical atmosphere, above and below that
approximately the same fluxes of muons are generated. For $\theta < 70^\circ$ we have used value $H$=15 km and
for big angles ($70^\circ - 90^\circ$) the approximation obtained in Ref.~\cite{BN} by using data tables of Ref.~\cite{ZK}:
\begin{equation}\label{Efcos3}
H=34- 10.5\cos\theta-120\cos^{2}\theta+250\cos^{3}\theta \qquad (\text{km}). 
\end{equation}
Since for $\xi>3.9$ the value of $(1-\text{erf}(\xi))$ is equal to $0$ with a good accuracy 
one should develop the expression $\exp (\xi^2)(1-\text{erf}(\xi))$ as a series in $\xi$ and rewrite 
~(\ref{Efcos1}) as:
\begin{equation}\label{Efcos4}
\cos\theta^{*}=\xi \sqrt{\frac {2H_0}{R_{E}+H}} \biggl 
( 1+ \sum_{n=1}^{\infty} \frac{(2n-1)!!}{(-2\xi^2)^n} \biggr )^{-1}.
\end{equation}
We have used~(\ref{Efcos4}) for $\cos\theta>\,$0.1 (that corresponds to $\xi>\,$3.1429) and ~(\ref{Efcos1}) for $\cos\theta$=(0--0.1).
In spite of rather slow convergence of series in~(\ref{Efcos4}), seven its terms are enough for accuracy of 0.1$\,\%$.          
For $\cos\theta>\,$0.8 the atmosphere can be considered as flat one (the deviation is less then 0.3$\,\%$). 
For an estimation avoiding the special function and series calculations we propose the polynomial 
fit of~(\ref{Efcos1}):  
\begin{equation}\label{Efcosfit}
\cos\theta^{*}=\sum_{i=0}^4 c_{i}\cos^{i}\theta,  
\end{equation}
with the coefficients of the decomposition assembled in Table~\ref{tab:ectab}.
\begin{table}[htb]
\protect\caption{Coefficients $c_{i}$ of the fitting formula~(\protect\ref{Efcosfit})
                 for effective cosine with the maximum relative errors.
\label{tab:ectab}}
\center{\begin{tabular}{ccccccc}
$\cos \theta$ & $c_0$  & $c_1$ & $c_2$ & $c_3$ & $c_4$ & Max.err,$\%$ \\\hline
0$\div$0.002    & 0.11137 & 0 & 0 & 0 & 0 & 0.004\\ 
0.002$\div$0.2  & 0.11148 & $-0.03427$ & 5.2053  & $-14.197$ & 16.138 & 0.3\\ 
0.2$\div$0.8    & 0.06714 & 0.71578  & 0.42377 & $-0.19634$ & $-0.021145$ & 0.7\\ 
\end{tabular}}
\end{table} 
The accuracy of~(\ref{Efcosfit}) is much better than 0.3$\,\%$ except the region $\cos\theta$=(0.3--0.38) where
it may reach the value of 0.7$\,\%$.  

\section{Derivation of formula for integral spectrum in matter}
\label{app:a2}
The solution of~(\ref{Ra}) in the assumption of continuous linear energy losses $L(E)=\alpha + \beta E$ is as follows: 
\begin{equation}\label{f1}
E_{s}(E_{f},R)=\frac{1}{\beta}((\alpha + \beta E_{f})e^{\beta R} - \alpha).  
\end{equation}
Here $E_s$ is that value of the starting energy $E_0$ of the muon at sea level which results to
passing of path $R$ in matter with the final energy $E_f$. 
Thus the expression~(\ref{Int_cont}) for the integral flux in matter (after change of designation $I_{cl} \to F_{cl}$)
is transformed to
\begin{equation}\label{f2}
F_{cl}(\geq E_f,R,\theta)=\int_{E_f}^{\infty}{D(E_{s}(E_{0},R),\theta) e^{\beta R}\,dE_{0}}.
\end{equation}
For simplicity let us keep only one term of series~(\ref{Sealds1}) for the sea level
differential spectrum, 
\begin{equation}\label{f3}
D(E_{0},\theta)=\frac {D_{0}E_{0}^{-\gamma}}{1+E_{0}/E^{cr}_{0}(\theta)},\qquad  
E^{cr}_{0}(\theta)=\frac {E^{cr}_{0}(0^\circ)}{\cos\theta^*}.
\end{equation}
After substitution of~(\ref{f3}) in~(\ref{f2}) one can show that the resulting expression
is reduced to 
\begin{equation}\label{f4}
F_{cl}(\geq E_f,R,\theta)=D_{0}E^{cr}_{0}(\theta) e^{-\beta R\gamma} 
\int_{E_f}^{\infty}{(E_{0}+x)^{-\gamma}(E_{0}+y)^{-1}\,dE_{0}},
\end{equation}
where $x=\alpha(1-e^{-\beta R})/\beta$ and $y=x+E^{cr}_{0}(\theta) e^{-\beta R}$.\\
Integral in~(\ref{f4}) has the tabulated solution (Ref.~\cite{Grad}) resulting in the following analytical expression
for integral flux in matter: 
\begin{equation}\label{f5}
F_{cl}(\geq E_f,R,\theta)=D_{0}E^{cr}_{0}(\theta) e^{-\beta R\gamma} (E_f+y)^{-\gamma} 
\text {B}(\gamma,1)\,{_2\text{F}_1}(\gamma,\gamma,\gamma+1,z), 
\end{equation}
where B$(\gamma,1)$ is Bessel function and ${_2\text{F}_1}(\gamma,\gamma,\gamma+1,z)$ is hypergeometric function 
of variable $z$ defined by the expression 
\[
z=E^{cr}_{0}(\theta) e^{-\beta R}/(E_{f}+y).
\]
All the conditions of convergence of ~(\ref{f5}) are fulfiled (let us omit here the proof because of long calculations). 
The expressions for special functions in~(\ref{f5}) can be reduced to
\[
\text {B}(\gamma,1)=\frac{1}{\gamma}, \qquad
{_2\text{F}_1}(\gamma,\gamma,\gamma+1,z)=(1-z)^{1-\gamma} \, \text{S}(z,\gamma),
\]
where
\begin{eqnarray}\label{f6}
\text{S}(z,\gamma)={_2\text{F}_1}(1,1,\gamma+1,z)= 1+\frac{z}{\gamma+1}+\frac{2z^2}{(\gamma+1)(\gamma+2)}+ 
\frac{6z^3}{(\gamma+1)(\gamma+2)(\gamma+3)} + \dots  \nonumber \\ 
= 1+ \sum_{n=1}^{\infty} n!\,z^{n} \biggl (\prod_{j=1}^{n} (\gamma+j) \biggr )^{-1}. 
\end{eqnarray}

Finally, taking into account both ($\pi$,K)-terms of the sea level spectrum~(\ref{Sealds1})
the expression for the integral angular flux in matter in the assumption of continuous losses is as follows: 
\begin{equation}\label{f7}
F_{cl}(\geq E_f,R,\theta)=\frac{e^{-\beta R\gamma}}{\gamma}\sum_{i={\pi,K}} D_{0_i} E^{cr}_{0_i}(\theta) 
(E_f+y_i)^{-\gamma} (1-z_i)^{1-\gamma}\,\text{S}(z_{i},\gamma). 
\end{equation}
In the case of a flat surface the slant depth is expressed by $R=h/\cos\theta$, where $h$
is a vertical depth below the surface.

It is important to note that formula(~\ref{f7}) evaluated in the assumption of the single slope $(\alpha,\beta)$ of linear total losses  
$\alpha+\beta E$ may be easily extended to the case of multislope losses. For example, in the case of a change of loss 
slope from $(\alpha_1,\beta_1)$ 
to $(\alpha_2,\beta_2)$ on the way of a muon along the path $R$ the integral equation~(\ref{Ra}) is transformed to   
\begin{equation}\label{Ram}
\int_{E_f}^{E_s} \frac{dE}{\alpha+\beta E}=\int_{E_f}^{E_{12}} \frac{dE}{\alpha_{1}+\beta_{1}E} + 
\int_{E_{12}}^{E_s} \frac{dE}{\alpha_{2}+\beta_{2}E}=R,
\end{equation}
where $E_{12}$ is energy value in the point of a slope change. Thus, to compute~(\ref{Int_cont})
one must determine $E_s$ from the integral equation 
\begin{equation}\label{Ram1}
\int_{E_{12}}^{E_s} \frac{dE}{\alpha_{2}+\beta_{2}E}=R-R_{12},
\end{equation}
where 
\begin{equation}\label{Ram2}
R_{12}=\frac {1}{\beta_1}\,\ln \biggl( \frac {\alpha_{1}+E_{12}\beta_{1}}{\alpha_{1}+E_{f}\beta_1} \biggr).
\end{equation}
It leads to the integral flux expressed by $F_{cl}(\geq E_f,R,\theta)=F_{cl}(\geq E_{12},(R-R_{12}),\theta;~\alpha_2,\beta_2)$.  
In other words, when using expression~(\ref{f7}) for the slant depths $R>R_{12}$ one must substitute $R \to (R-R_{12})$ 
and $E_f \to E_{12}$ and use the values ($\alpha_2,\beta_2$) for a loss description. For slant depths $R \leq R_{12}$ the use of~
(\ref{f7}) remains unchangeable and the loss values are expressed by ($\alpha_1,\beta_1$).

This algorithm may be extended to computations with any number of slopes of the energy dependence $L(E)$.


\newpage

\end{document}